\documentclass[a4paper,fleqn,usenatbib]{mnras}
\usepackage[T1]{fontenc}
\usepackage{ae,aecompl}

\usepackage{amsmath}
\usepackage{epsfig}
\usepackage{graphicx}
\usepackage{lscape}
\usepackage{subfig}
\usepackage{amssymb}
\usepackage{color}
\usepackage{url}
\newcommand {\bc}{\begin {center}}
\newcommand {\ec}{\end {center}}
\newcommand {\be}{\begin {equation}}
\newcommand {\ee}{\end {equation}}
\newcommand {\beq}{\begin {eqnarray}}
\newcommand {\eeq}{\end {eqnarray}}
\newcommand {\ergs}{{\rm erg\ \rm s$^{-1}$}}

\def\lum{erg s$^{-1}$}

\title[The X-ray properties of Be/XRPs in quiescence]
{The X-ray properties of Be/X-ray pulsars in quiescence}

\author[S.~S.~Tsygankov et al.]
{Sergey~S.~Tsygankov,$^{1}$\thanks{E-mail: stsygankov@gmail.com}
Rudy~Wijnands,$^{2}$
Alexander A.~Lutovinov,$^{3,4}$
Nathalie~Degenaar,$^{5}$ \newauthor
and Juri~Poutanen$^{1,6,7}$
 \\
$^1$ Tuorla Observatory, Department of Physics and Astronomy, University of Turku,
  V\"ais\"al\"antie 20, FI-21500 Piikki\"o, Finland \\
$^2$ Anton Pannekoek Institute of Astronomy, University of Amsterdam, Science Park 904, 1098 XH Amsterdam, The Netherlands \\
$^3$ Space Research Institute of the Russian Academy of Sciences, Profsoyuznaya Str. 84/32, Moscow 117997, Russia \\
$^4$ Moscow Institute of Physics and Technology, Moscow region, Dolgoprudnyi 141701, Russia \\
$^5$ Institute of Astronomy, University of Cambridge, Madingley Road, Cambridge CB3 OHA, UK \\
$^6$ Nordita, KTH Royal Institute of Technology and Stockholm University, Roslagstullsbacken 23, SE-10691 Stockholm, Sweden\\
$^7$ Kavli Institute for Theoretical Physics, University of California, Santa Barbara, CA 93106, USA
}

\date{Accepted 2017 May 17. Received 2017 May 17; in original form 2017 March 14}

\pubyear{2017}

\begin{document}
\label{firstpage}
\pagerange{\pageref{firstpage}--\pageref{lastpage}}
\maketitle

%%%%%%%%%%%%%%%%%%%%%%%%%%%%%%%%%%%%%%%%%%%%%%%%%%%%%%%%%%%%%%%%%%%%%%%%%%%%%%
%% Abstract, Keywords and contact details                                   %%
%%%%%%%%%%%%%%%%%%%%%%%%%%%%%%%%%%%%%%%%%%%%%%%%%%%%%%%%%%%%%%%%%%%%%%%%%%%%%%
\begin{abstract}
Observations of accreting neutron stars (NSs) with strong magnetic fields can
be used not only for studying the accretion flow interaction with the NS
magnetospheres, but also for understanding the physical processes inside
NSs and for estimating their fundamental parameters. Of particular
interest are (i) the interaction of a rotating NS (magnetosphere)
with the infalling matter at different accretion rates, and (ii) the theory
of deep crustal heating and the influence of a strong magnetic field on this
process. Here, we present results of the first systematic investigation of 16
X-ray pulsars with Be optical companions during their quiescent states, based
on data from the {\it Chandra}, {\it XMM--Newton} and {\it Swift}
observatories. The whole sample of sources can be roughly divided into two
distinct groups: (i) relatively bright objects with a luminosity around
$\sim10^{34}$\,\lum\ and (hard) power-law spectra, and (ii) fainter ones
showing thermal spectra. X-ray pulsations were detected from five objects in
group (i) with quite a large pulse fraction of 50--70 per cent. The obtained results
are discussed within the framework of the models describing the interaction
of the infalling matter with the NS magnetic field and those
describing heating and cooling in accreting NSs.
\end{abstract}


%%%%%%%%%%%%%%%%%%%%%%%%%%%%%%%%%%%%%%%%%%%%%%%%%%%%%%%%%%%%%%%%%%%%%%%%%%%%%%
%% Introduction                                                             %%
%%%%%%%%%%%%%%%%%%%%%%%%%%%%%%%%%%%%%%%%%%%%%%%%%%%%%%%%%%%%%%%%%%%%%%%%%%%%%%
\section{Introduction}
\label{intro}

Transient X-ray pulsars in binary systems with Be optical companions
(BeXRPs) are unique laboratories that allow us to test different
theories of the accretion on to the magnetized neutron stars (NSs).  This is
possible due to an extremely broad dynamical range of X-ray
luminosities demonstrated by such systems: from $\sim10^{32-33}$ \lum\ in
quiescence to more than $10^{38-39}$ \lum\ during powerful outbursts.

In general, BeXRPs manifest themselves mainly through transient
activity of two types: type I -- periodically appearing outbursts
(once per binary orbit, during the periastron passage) with an X-ray
luminosity of $\lesssim10^{37}$ \lum\ and duration of 20--30 per cent  (i.e. typically about one week) of
the orbital period; and type II -- giant and
rare outbursts (usually once in several years) with the peak
luminosity exceeding several times $10^{37}-10^{38}$ \lum\ and lasting
from several weeks to months. A review of the observational properties
of BeXRPs can be found in \citet{2011Ap&SS.332....1R}.

The presence of a strong magnetic field in X-ray pulsars determines the main
observational properties of these objects.  For instance, the maximum
observed luminosity may exceed the Eddington limit due to the appearance of
accretion columns at the magnetic poles of the NSs
\citep{1976MNRAS.175..395B}, making them among the brightest accreting objects
on the sky \citep[see e.g. recent works by][and references
therein]{2014Natur.514..202B,2015MNRAS.454.2539M,2017Sci...355.817I}. Therefore, the
temporal and spectral properties of such systems are very sensitive to the
magnetic field strength.

%=================================================================
%\begin{landscape}
\begin{table*}
\noindent
\centering
\caption{Sample of the studied BeXRPs.}\label{tabsam}
\centering
\vspace{1mm}
%\small{
\begin{tabular}{|l|c|c|c|c|c|}
\hline\hline
   Source name           & Pulse  period    & Orbital  period  & Distance & Cyclotron line & Optical  companion   \\
                & (s) & (d) & (kpc)       &  (keV) &    \\
\hline
4U\,0115+63          & 3.6 & 24.3$^{1}$ & 7.0$^{2}$  & 11.5$^{3}$ & B0.2Ve$^{2}$  \\
V\,0332+53           & 4.375 & 34.25$^{4}$ & 7.0$^{5}$  & 28$^{6}$   &  O8--9Ve$^{5}$ \\
MXB\,0656$-$072        & 160.4 & 101.2$^{7}$ & 3.9$^{8}$  &  36$^{9}$  & O9.7Ve$^{10}$  \\
4U\,0728$-$25          & 103 & 34.5$^{11}$ &  6.1$^{12}$    &  --  & O8--9Ve$^{12}$  \\
RX\,J0812.4$-$3114     & 31.9 & 81.3$^{13}$ &  9.2$^{14}$   & --  & B0.5V-IIIe$^{14}$  \\
GS\,0834$-$430         & 12.3 & 105.8$^{15}$ & 5.0$^{16}$  &  --  & B0--2 V--IIIe$^{16}$  \\
GRO\,J1008$-$57        & 93.8 &  249.5$^{17}$ &  5.8$^{18}$  & 88$^{19}$, 75.5$^{20}$  &  B0e$^{21}$ \\
2S\,1417$-$624         & 17.6 & 42.1$^{22}$ &  (1.4--11.1)$^{23}$   & --   &  B1 Ve$^{23}$ \\
2S\,1553$-$542         & 9.3 & 30.6$^{24}$ &   20$^{25,26}$   & 23.5$^{25}$  & --  \\
Swift\,J1626.6$-$5156  & 15.377 & 132.9$^{27}$ & 10$^{28}$  & 10$^{29}$   & B0Ve$^{28}$  \\
GS\,1843+00          & 29.5 & -- &  (10--15)$^{30}$ & 20$^{31}$  & B0--B2IV--Ve$^{30}$  \\
XTE\,J1946+274       & 15.8 & 169.2$^{32}$ & (8--10)$^{33}$ & 36$^{34}$  & B0--B1 IV--Ve$^{33}$  \\
KS\,1947+300         & 18.8 & 40.4$^{35}$ &  10$^{36,37}$ & 12.5$^{38}$   & B0Ve$^{36}$  \\
SAX\,J2103.5+4545    & 351 & 12.7$^{39}$ &  6.5$^{40}$  &  --  & B0Ve$^{40}$  \\
Cep\,X-4             & 66.3 & 21$^{41}$ & 3.8$^{42}$  & 30$^{43}$  & B1V--B2Ve$^{42}$  \\
SAX\,J2239.3+6116    & 1247 & 262$^{44}$ & 4.4$^{44}$  & --   & (B0--2 V--IIIe)$^{44}$  \\
\hline
\end{tabular}
\vspace{3mm}
\begin{minipage}{0.8\textwidth}
(1) \cite{1978Natur.273..367C}, (2) \cite{2001A&A...369..108N}, (3) \cite{1983ApJ...270..711W}, (4) \cite{1985ApJ...288L..45S}, (5) \cite{1999MNRAS.307..695N}, (6) \cite{1990ApJ...365L..59M}, (7) \cite{2012ApJ...753...73Y}, (8) \cite{2006A&A...451..267M}, (9) \cite{2003ATel..200....1H}, (10) \cite{2003ATel..202....1P}, (11) \cite{1997ApJ...489L..83C}, (12) \cite{1996A&A...315..160N}, (13) \cite{2000ApJ...530L..33C}, (14) \cite{1997A&A...323..853M}, (15) \cite{1997ApJ...479..388W}, (16) \cite{2000MNRAS.314...87I}, (17) \cite{2012ATel.4564....1K}, (18) \cite{2012A&A...539A.114R}, (19) \cite{1999ApJ...512..920S}, (20) \cite{2013ATel.4759....1Y}, (21) \cite{1994MNRAS.270L..57C}, (22) \cite{1996A&AS..120C.209F}, (23) \cite{1984ApJ...276..621G}, (24) \cite{1983ApJ...274..765K}, (25) \cite{2016MNRAS.457..258T}, (26) \cite{2016MNRAS.462.3823L}, (27) \cite{2010ApJ...711.1306B}, (28) \cite{2011A&A...533A..23R}, (29) \cite{2013ApJ...762...61D}, (30) \cite{2001A&A...371.1018I}, (31) \cite{1995PhDT.......215M}, (32) \cite{2003ApJ...584..996W}, (33) \cite{2002A&A...393..983V}, (34) \cite{2001ApJ...563L..35H}, (35) \cite{2004ApJ...613.1164G}, (36) \cite{2003A&A...397..739N}, (37)
\cite{2005AstL...31...88T}, (38) \cite{2014ApJ...784L..40F}, (39) \cite{2000ApJ...544L.129B}, (40) \cite{2004A&A...421..673R}, (41) \cite{2007MNRAS.382..743M}, (42) \cite{1998A&A...332L...9B}, (43) \cite{1991ApJ...379L..61M}, (44) \cite{2000A&A...361...85I}.
\end{minipage}
\end{table*}
%\end{landscape}
%=================================================================

BeXRPs can act as unique laboratories thanks to the quite accurate knowledge
of the magnetic field strength of the NSs in a substantial fraction
of such systems. This has been achieved through the detection of cyclotron
absorption features in the spectra of X-ray pulsars and accurate measurements of
their energies. Such features are known in more than two dozens
of sources \citep[see the recent review of high-mass X-ray binaries
by][]{2015A&ARv..23....2W}. Measured values of the cyclotron energy vary between $\sim5$
and $\sim80$ keV, which correspond to the magnetic field strengths on the NS
surface ranging from $\sim5\times10^{11}$ to
$\sim8\times10^{12}$ G. Typical values of spin periods in such systems
range from seconds to hundreds of seconds.

In turn, such strong fields can support a magnetosphere  of the
order of $10^8$--$10^9$ cm, which in some cases can be significantly
larger than the corotation radius. In this case one can expect a
termination of the accretion flow due to the centrifugal forces -- the
so-called propeller regime
\citep{1975A&A....39..185I,1985ApJ...288L..45S}. At the same time,
observations of some BeXRPs in a quiescence contradict these
expectations.  Particularly, a few sources demonstrate X-ray
pulsations at luminosities of $10^{33}$ -- $10^{34}$
\ergs\ \citep{1991ApJ...369..490M,2000A&A...356.1003N,
  2004NuPhS.132..476O,2007ApJ...658..514R,2013ApJ...770...19R,
  2014A&A...561A..96D,2014MNRAS.445.1314R}. In different systems the
nature of these pulsations is likely different. For pulsars
with large spin periods (dozens of seconds) the required leakage of matter through the
magnetospheric barrier  can be achieved via quasi-stable accretion from
a cold recombined disc \citep{tsyg2017}.

Another possible origin of the NS emission in quiescent
BeXRPs is their thermal energy accumulated during accretion episodes.
\citet{1998ApJ...504L..95B} proposed a model in which the NS
core is maintained at temperatures of about $10^8$ K by nuclear
reactions in the crust, that occur when it is compressed by freshly
accreted material; so-called deep crustal heating. The resulting
thermal emission from the NS at its surface is proportional to the
average (over thousands to tens of thousands of years) mass accretion
rate $\langle \dot{M} \rangle$ and can be as high as
$\sim6\times10^{32}$ \ergs\ for $\langle \dot{M}
\rangle\simeq10^{-11}$ M$_{\odot}$ yr$^{-1}$. This theory is well
supported by observations of a number of X-ray binaries (i.e. those
harbouring a low magnetic field NS in low-mass X-ray
binaries; LMXBs) in quiescence, although for many systems enhanced
neutrino cooling processes in the NS core have to be assumed to
explain their low temperatures \citep[see
  e.g.][]{1998ApJ...504L..95B,2004ARA&A..42..169Y,
  2009ApJ...691.1035H, 2010ApJ...714..894H, 2013MNRAS.432.2366W}.

Strong magnetic fields may affect the NS
heating and cooling processes during the accretion episodes and between them.
In contrast to low magnetic field NSs in LMXBs the accretion in X-ray pulsars proceeds
to a fraction of the NS surface as small as $\sim0.01$ per cent
\citep{2015MNRAS.447.1847M}.
It is currently unclear how the heating of the NS is affected by such
inhomogeneous accretion process. Furthermore, the original NS crust in BeXRPs
might not have been fully replaced yet with accreted matter (a so-called
hybrid crust) that could cause some of the deep crustal heating reactions to
be altered or not even to occur \citep[see e.g.][]{2013MNRAS.432.2366W}. In
addition, the presence of a strong magnetic field could cause an anisotropy
of the thermal  conductivity in the crust and therefore likely an anisotropic
surface temperature distribution, that affects the cooling history of the
NS \citep[see e.g.][]{1996A&A...309..171S,2006A&A...457..937G,
2008ApJ...673L.167A}. At the moment no systematic studies of the cooling of
NSs with strong magnetic fields has been performed, and therefore
only limited information is available \citep[for a
discussion see][]{2001ApJ...561..924C, 2002ApJ...580..389C, 2013MNRAS.432.2366W,
2016MNRAS.463...78E}. BeXRPs are the best objects for such studies, because
they often exhibit distinct outburst and quiescence episodes.

The aim of this work is to study systematically the timing and spectral
properties of X-ray pulsars with Be optical companions in their quiescent
states. The obtained results give insights into the problem of the accretion
on to magnetized NSs at very low mass accretion rates and the
possibility to study cooling of the NSs in such systems.

\section{Data analysis and results}

\subsection{Observations}
\label{sec:obs}

This work is based mainly on the data from the {\it Chandra} observatory
acquired in the frame of a specially dedicated observational campaign to
study quiescent BeXRPs (PI R. Wijnands, ObsIDs. 14635-14650). Additionally we
used the publicly available data from the {\it Chandra} (ObsIDs. 1919, 10049)
and {\it XMM-Newton} (ObsIDs. 0505280101, 0506190101, 0302970201)
observatories, obtained for two sources -- 4U\,0115+63 and V\,0332+53,
as well as the data from the XRT telescope onboard the {\it
Swift} observatory, obtained for 4U\,0728$-$25 in the low state (ObsIDs.
00038005001--00038005004). The sample of 16 sources investigated in this work
is presented in Table\,\ref{tabsam} and contains the information about their
pulse and orbital periods, distances, optical companions and cyclotron line
energies with the corresponding references.  The magnetic field of the
NS can be estimated from the cyclotron energy using the simple
equation $B_{12}=(1+z) E_{\rm cyc}/11.6$ G, where $B_{12}$ the magnetic field
strength in units of 10$^{12}$ G, $z$ is the gravitational redshift and
$E_{\rm cyc}$ is the cyclotron line energy in keV. Note that there is a quite
large uncertainty for the distance estimations of several sources, especially
for 2S\,1417$-$624. In such cases we used the averaged values for the
luminosity calculations, i.e. 9 kpc for XTE\,1946+274, 12.5 kpc for
GS\,1843+00, and 6 kpc for 2S\,1417$-$624.

The {\it Chandra} observations were performed with the ACIS instrument with
typical exposures of 5 ks and the {\it XMM-Newton} observations with the EPIC
cameras, had exposures of about 25--30 ks. The data collected by the ACIS
instrument were reduced with the standard software package {\sc CIAO
4.7}\footnote{\url{http://cxc.harvard.edu/ciao/}} with CALDB v4.6.5. One data set
(ObsID. 1919) was taken very early in the {\it Chandra} mission, and
therefore we reprocessed it with the {\sc chandra\_repro} tools. We extracted
spectra and light curves of the sources from circular regions with 2.5--5\arcsec\
radii (depending on the source intensity). Background events were obtained
from circular regions (with radius of 30\arcsec) offset from the sources
position.  The source 4U\,0728$-$25 was relatively bright during its
observations with {\it Chandra}, which led to pile-up in the data.  It was
taken into account in the subsequent analysis according to the {\sc CIAO}
threads,\footnote{\url{http://cxc.harvard.edu/ciao/download/doc/pileup_abc.pdf}}
namely, by adding the {\sc pileup} model.

To process the {\it XMM-Newton} data, we used version 14.0 of the
\emph{XMM-Newton Science Analysis System (SAS)}. After the standard pipeline
processing, we searched for possible intervals of high background and
rejected them. This led to a decrease in the effective exposure time by
50-60 per cent down to $\simeq6-14$ ks (see details in Table\,\ref{tab:spec_all}).
For the analysis of the EPIC data we selected events with patterns in the
range 0--4 for the pn camera and 0--12 for the two MOS cameras, using a
circular region with a radius of 20\arcsec\ around the source positions.
Similarly to our {\it Chandra} analysis, background events were selected from
circular regions (with radius of 30\arcsec) offset from the source
positions.

The {\it Swift}/XRT telescope observed practically all our objects
many times, but the vast majority of these observations were performed
during outburst activity. We found that only observations of
4U\,0728--25 are suitable for the purpose of this paper. This source
was observed four times in 2008 (three times at the end of July and
once in December). The source flux during the July observations
(ObsIDs. 00038005001, 00038005002, 00038005003; all done in the Photon
Counting mode) was approximately constant, therefore we were able to
combine them to get a source spectrum with better statistics. The XRT
spectra were prepared using the online tools provided by the UK Swift
Science Data
Centre\footnote{\url{http://www.swift.ac.uk/user_objects/}}
\citep{2007A&A...469..379E,2009MNRAS.397.1177E}.

All observations were also inspected for the presence of coherent signals --
pulsations. Note that the time resolution for the {\it Chandra}/ACIS data is about
3.2 s, which is
insufficient to search for pulsations of short-period pulsars. Moreover the
measured count rates from most of our sources are very low and the exposures
quite short, which also limits the possibility to detect the pulsations.

To trace the long-term history of our sources and to estimate their average
accretion rates we used data from the {\it RXTE}/ASM \citep[the 2--10 keV
energy band]{1993A&AS...97..355B} and {\it Swift}/BAT monitors \citep[the
15--50 keV energy band]{2013ApJS..209...14K}. For illustration purposes in
Fig.\,\ref{fig:lcs}, we present light curves obtained using the {\it
Swift}/BAT all-sky monitor for all sources except RX\,J0812.4$-$3114 and
SAX\,J2239.3+6116, which were not detected in the past 20 years. The raw
light curves were downloaded from the {\it Swift}/BAT Hard X-ray Transient
Monitor
webpage{\footnote{\url{http://swift.gsfc.nasa.gov/results/transients/}} and
converted to mCrab units assuming a constant count rate from the Crab
nebula of 0.22 cts cm$^{-2}$ s$^{-1}$ in the 15--50 keV energy range.
Depending on the typical count rate from each source, an averaging interval of
one or five days was chosen.  In Fig.\,\ref{fig:lcs}, the times of the
observations used in this work are indicated by a horizontal line with the
name of the corresponding observatory. In the cases of 4U\,0115+63,
V\,0332+53, GS\,0834$-$43 and KS\,1947+300 the data from {\it
RXTE}/ASM{\footnote{\url{http://xte.mit.edu/ASM_lc.html}} are also shown with
grey points to illustrate the sources behaviour prior to the {\it Swift}/BAT
observations. Flux measurements were converted to mCrab units similar to
the {\it Swift}/BAT data assuming a constant count rate from the Crab nebula
of 75 cts s$^{-1}$.

%==============LCURVEs=======================================
\begin{figure*}
\centering

\hbox{
\includegraphics[height=0.23\textwidth,angle=0,bb=25 415 567 675,clip]{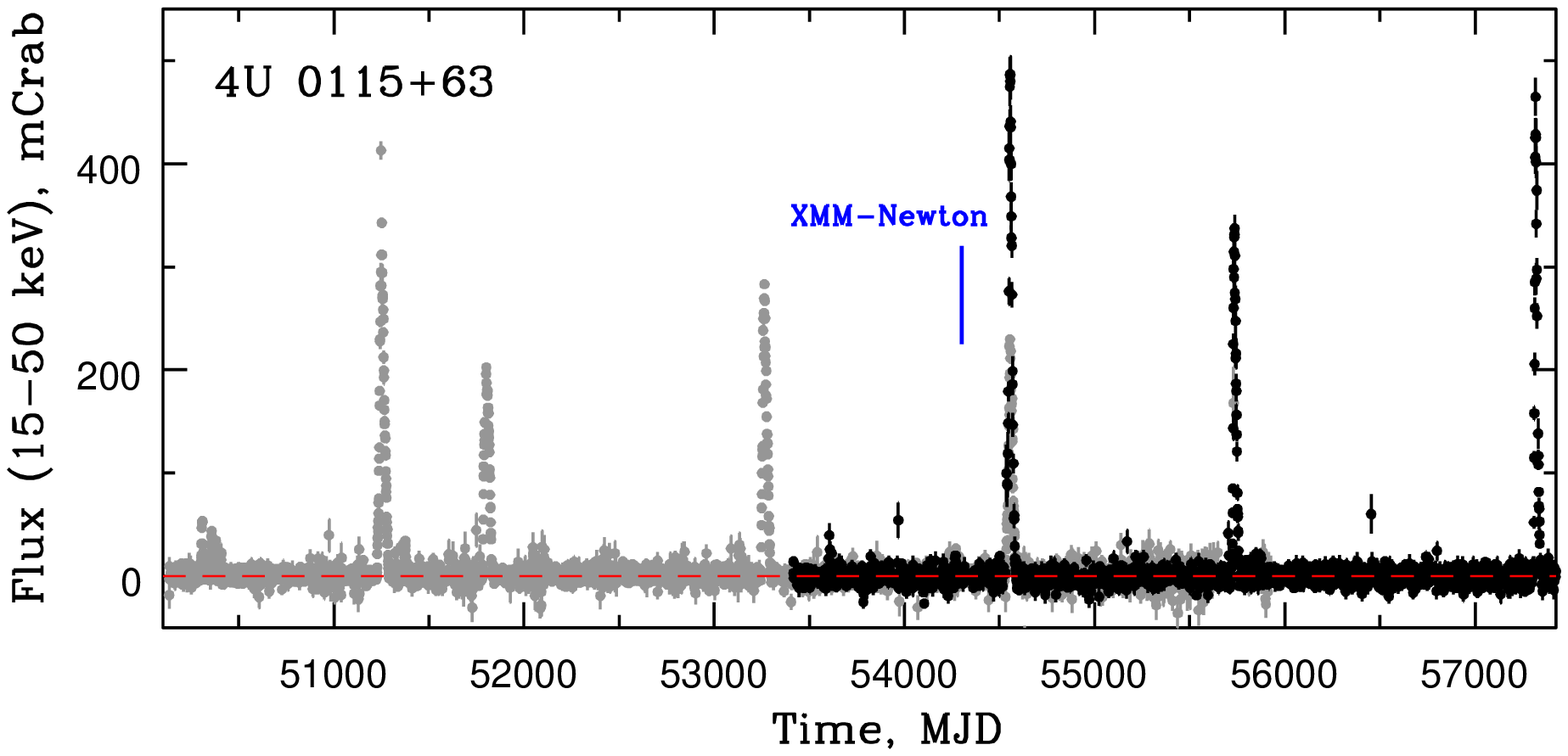}
\hspace{2mm}\includegraphics[height=0.23\textwidth,angle=0,bb=25 415 567 675,clip]{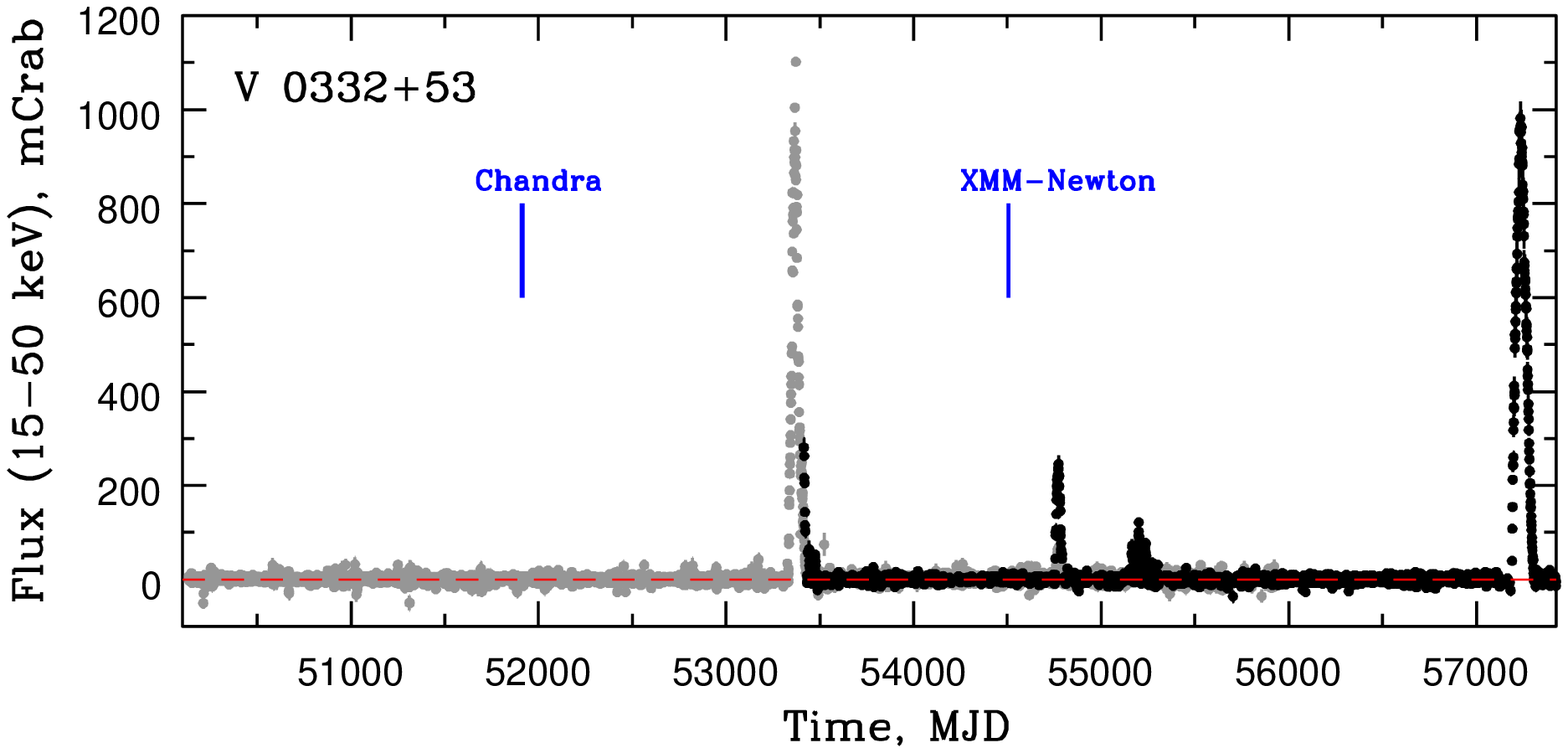}
}
\vspace{3mm}

\hbox{
\includegraphics[height=0.23\textwidth,angle=0,bb=25 415 567 675,clip]{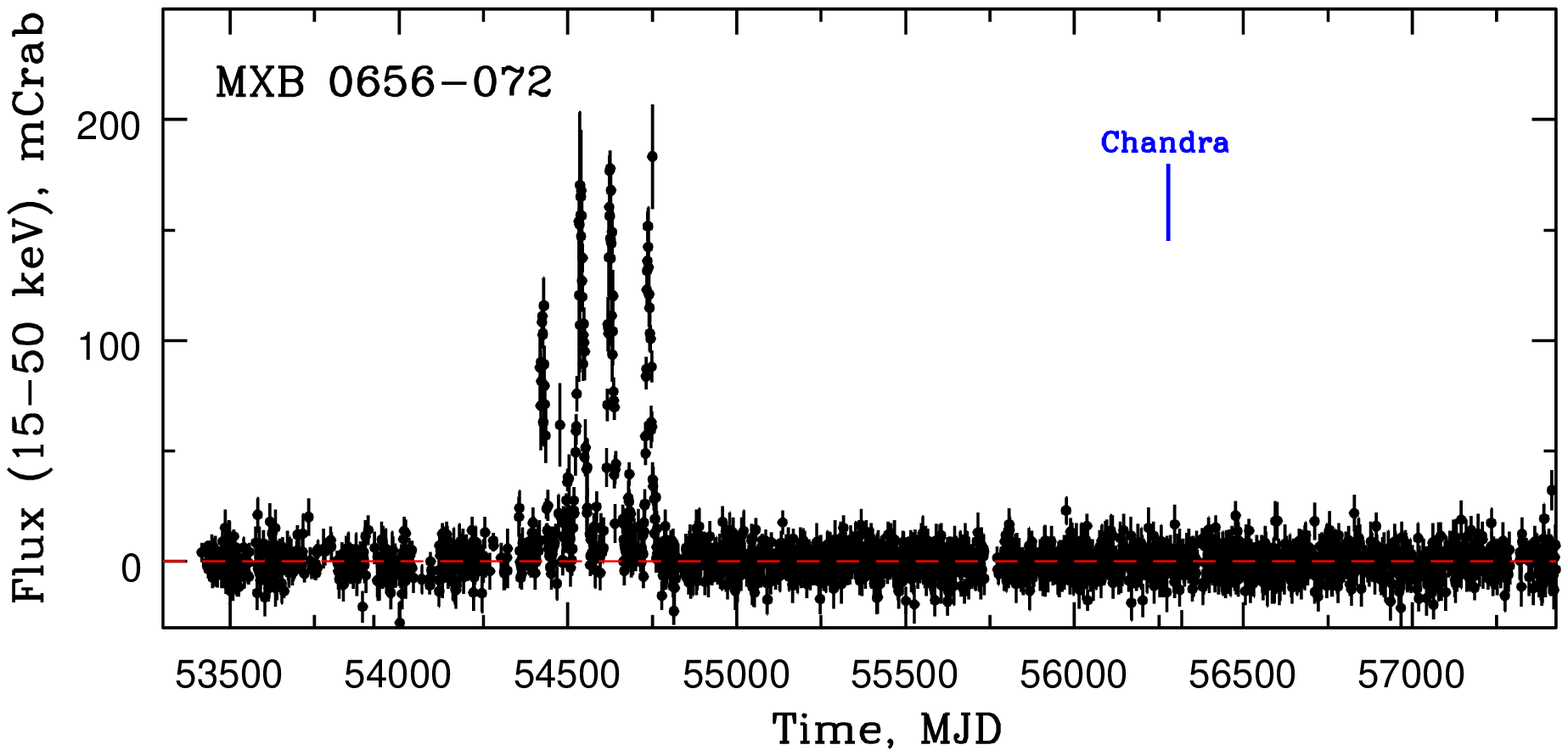}
\hspace{2mm}\includegraphics[height=0.23\textwidth,angle=0,bb=25 415 567 675,clip]{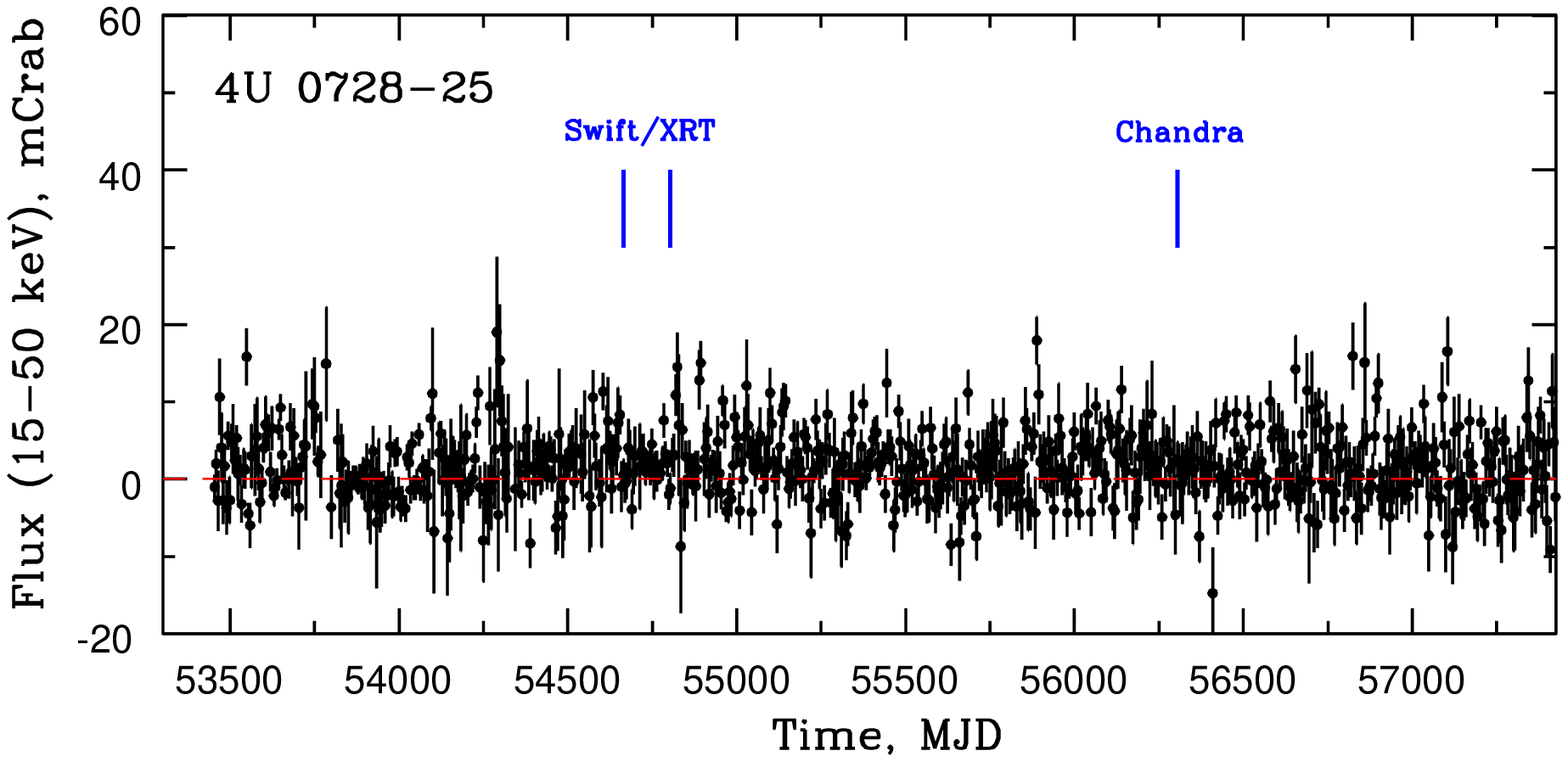}
}
\vspace{3mm}

\hbox{
\includegraphics[height=0.23\textwidth,angle=0,bb=25 415 567 675,clip]{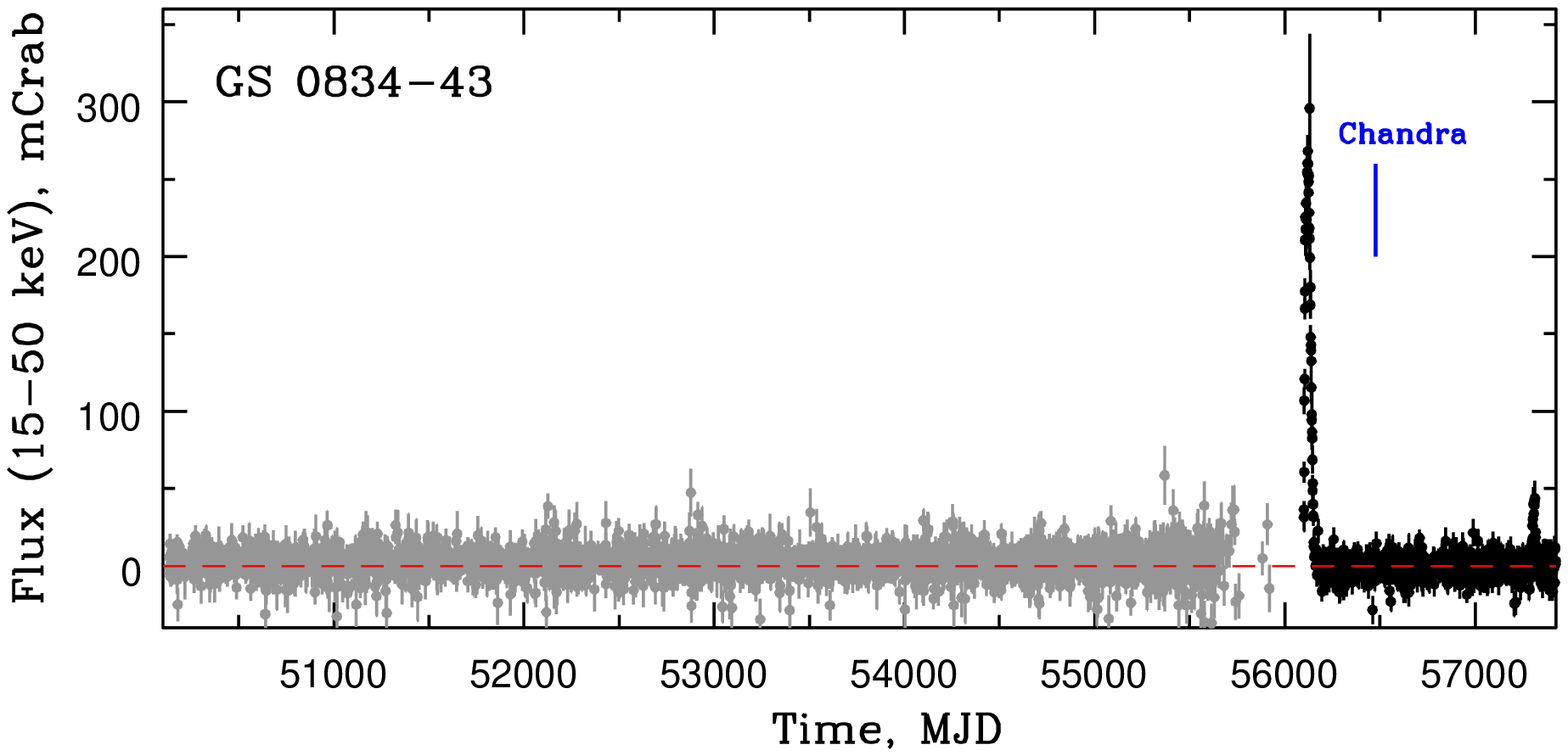}
\hspace{2mm}\includegraphics[height=0.23\textwidth,angle=0,bb=25 415 567 675,clip]{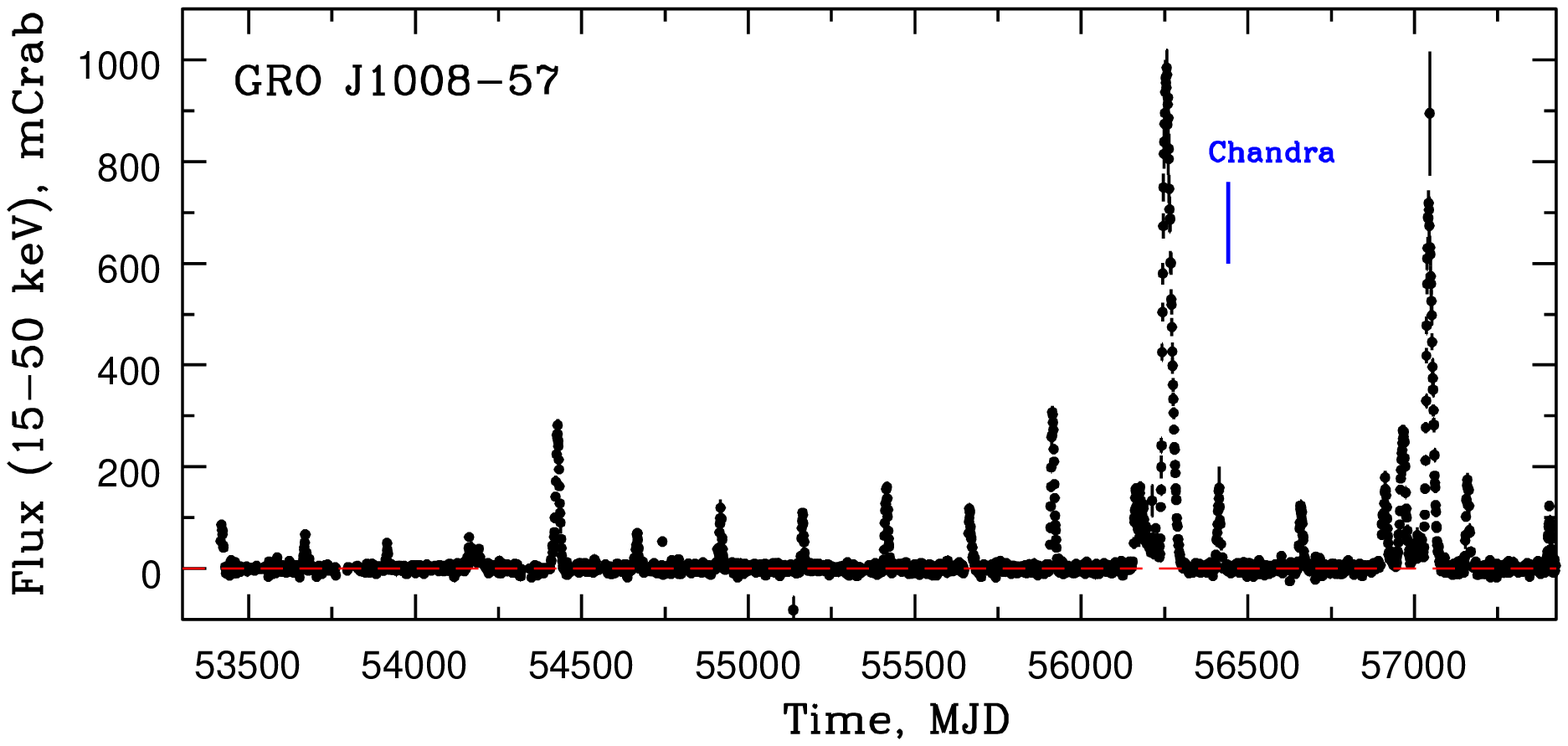}
}
\vspace{3mm}

\hbox{
\includegraphics[height=0.23\textwidth,angle=0,bb=25 415 567 675,clip]{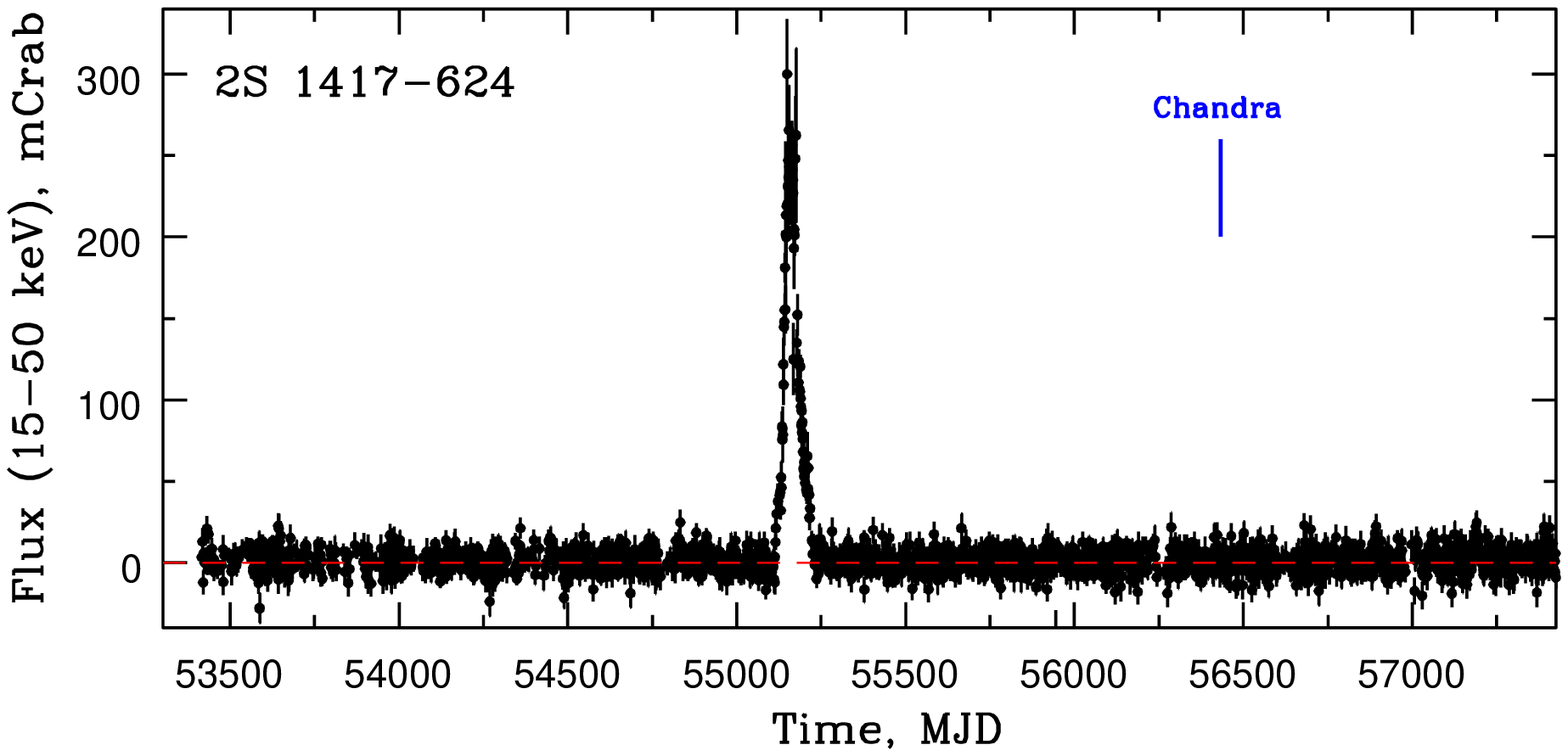}
\hspace{2mm}\includegraphics[height=0.23\textwidth,angle=0,bb=25 415 567 675,clip]{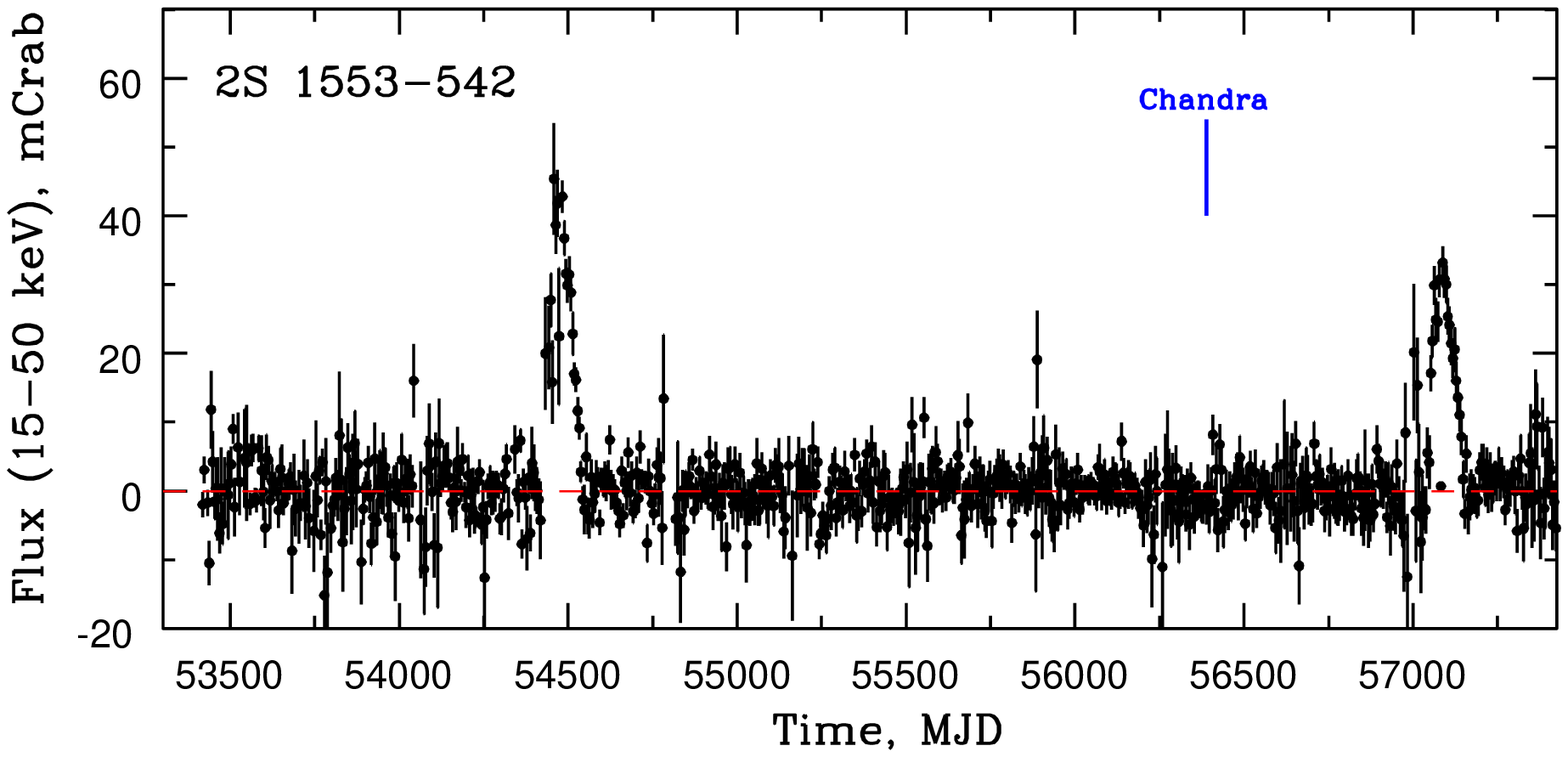}
}
\vspace{3mm}

\hbox{
\includegraphics[height=0.23\textwidth,angle=0,bb=25 415 567 675,clip]{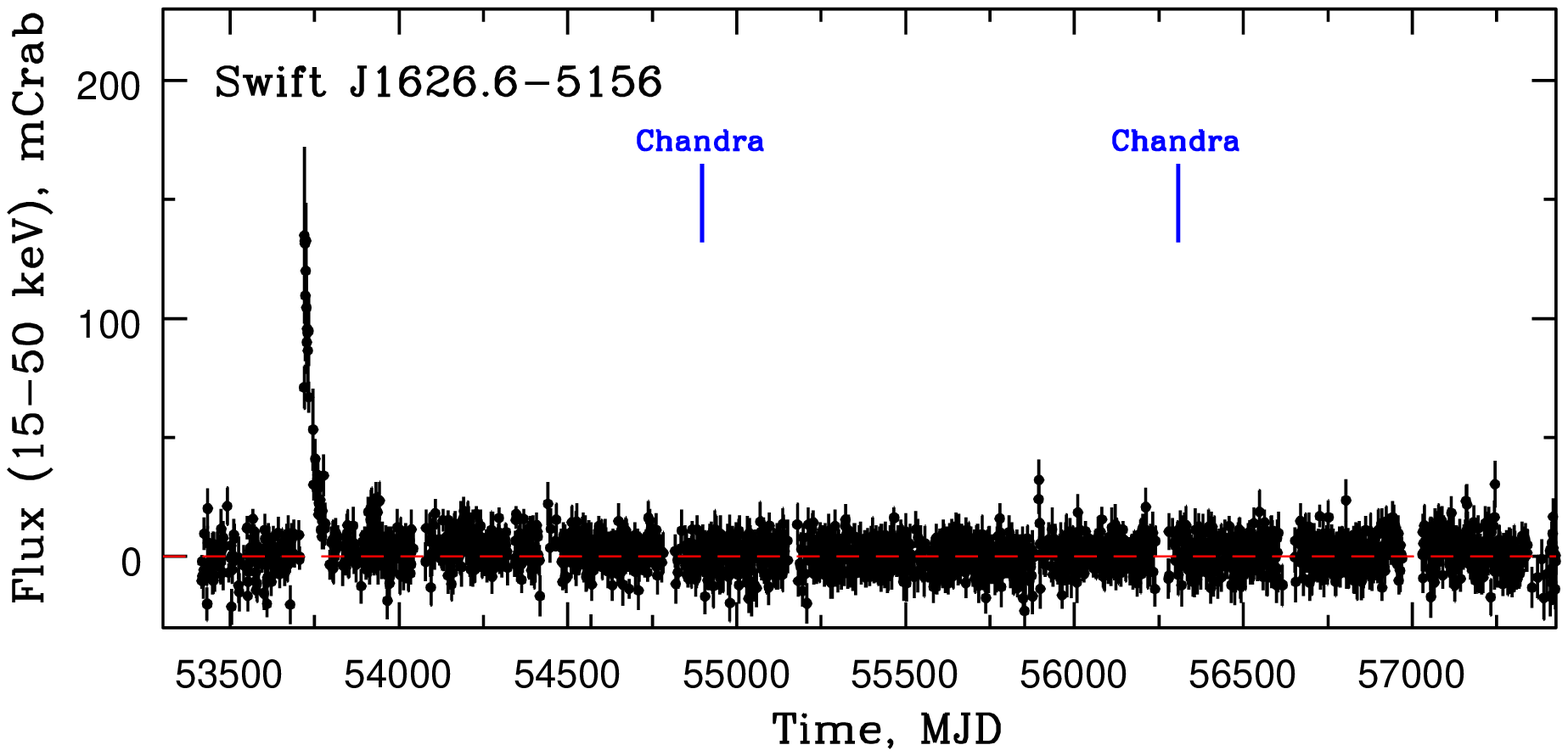}
\hspace{2mm}\includegraphics[height=0.23\textwidth,angle=0,bb=25 415 567 675,clip]{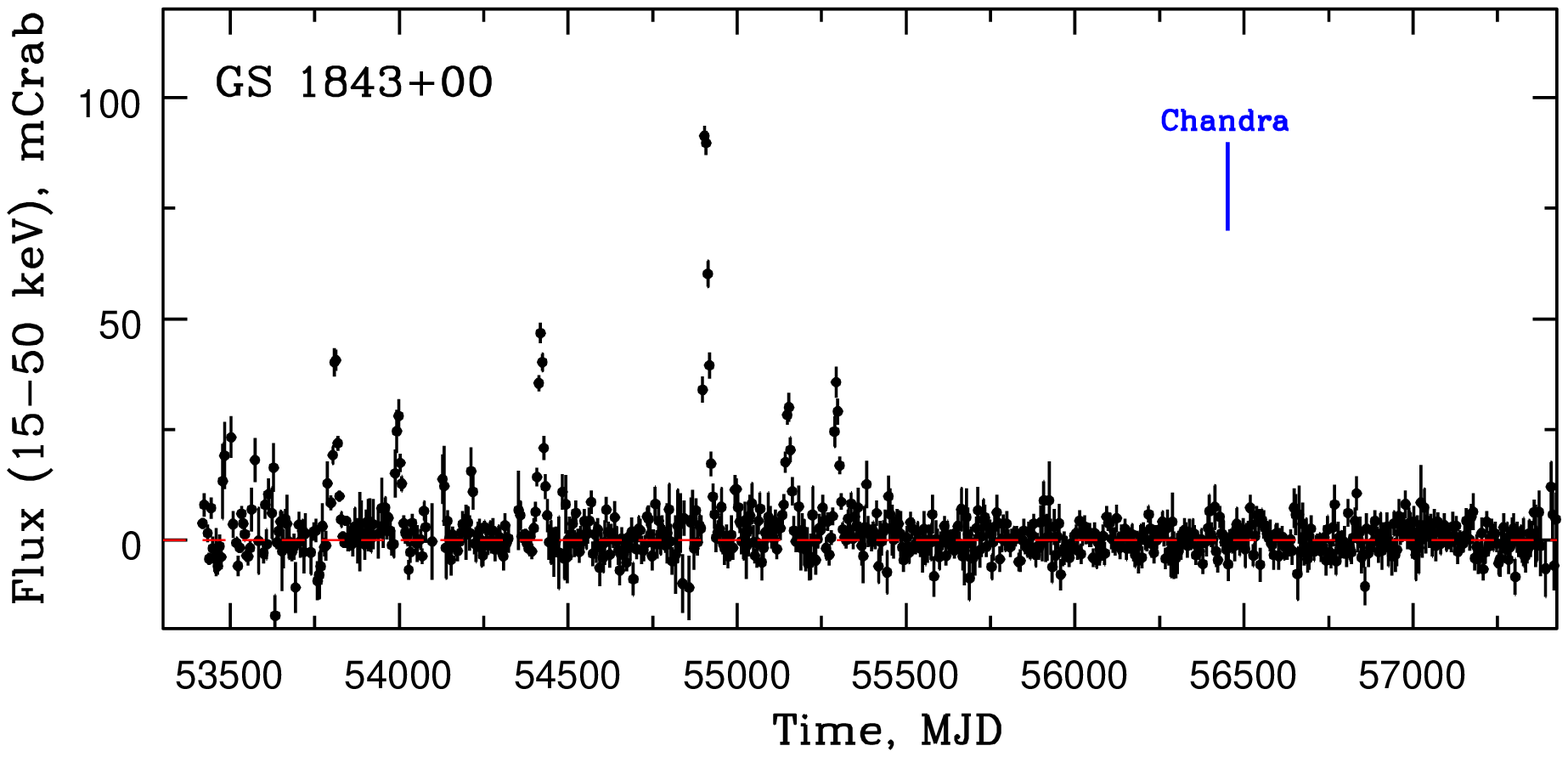}
}

\caption{{\it Swift}/BAT light curves in the 15--50 keV energy range. The
times of the {\it Chandra}, {\it XMM-Newton} and {\it Swift}/XRT observations
utilized in this work are indicated by the vertical bars. {\it RXTE}/ASM
light curves (grey points) in the 2--10 energy range are shown also for several sources.}\label{fig:lcs}
\end{figure*}
%=================================================================
%==============LCURVEs=======================================
\begin{figure*}
\centering

%\ContinuedFloat

\hbox{
\includegraphics[height=0.23\textwidth,angle=0,bb=25 415 567 675,clip]{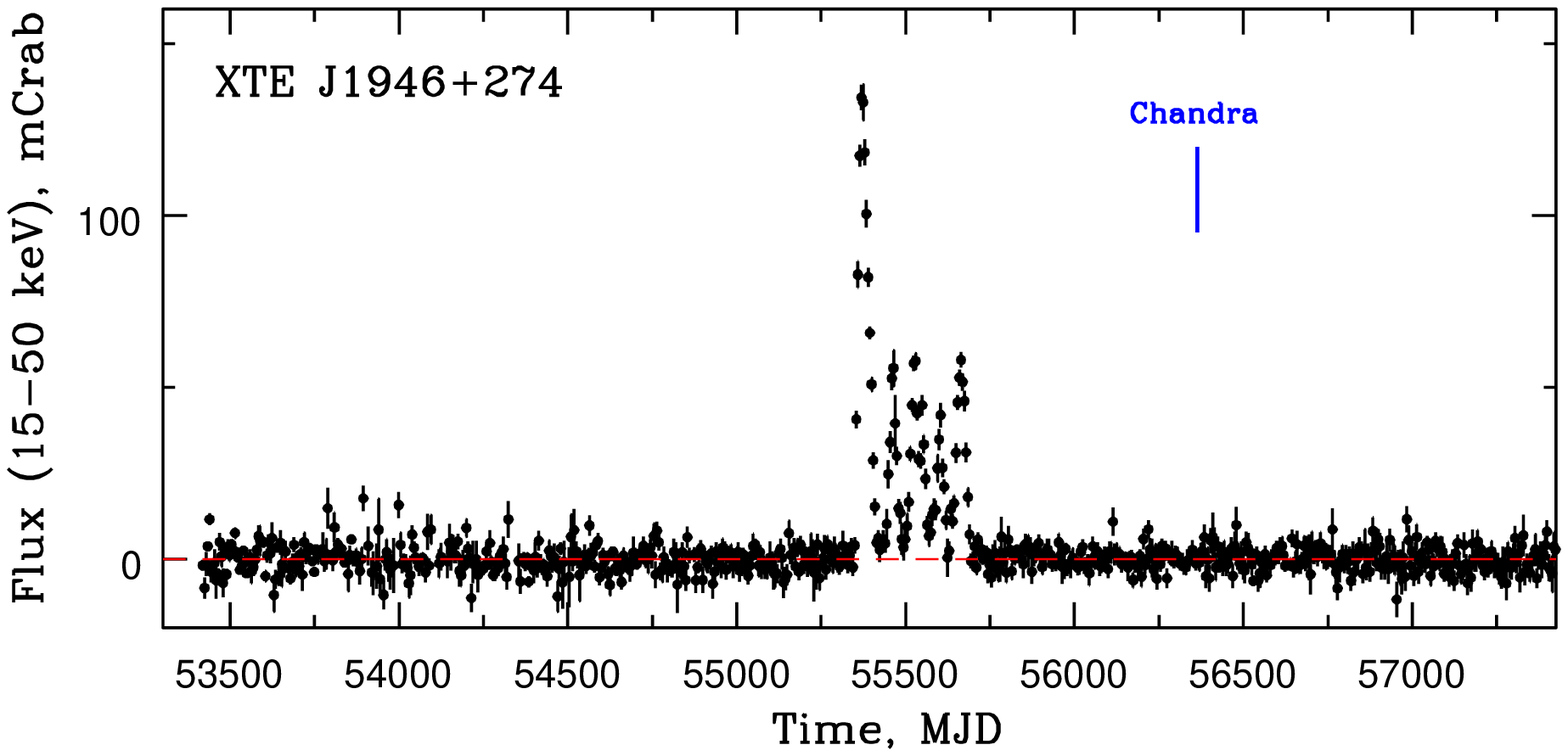}
\hspace{2mm}\includegraphics[height=0.23\textwidth,angle=0,bb=25 415 567 675,clip]{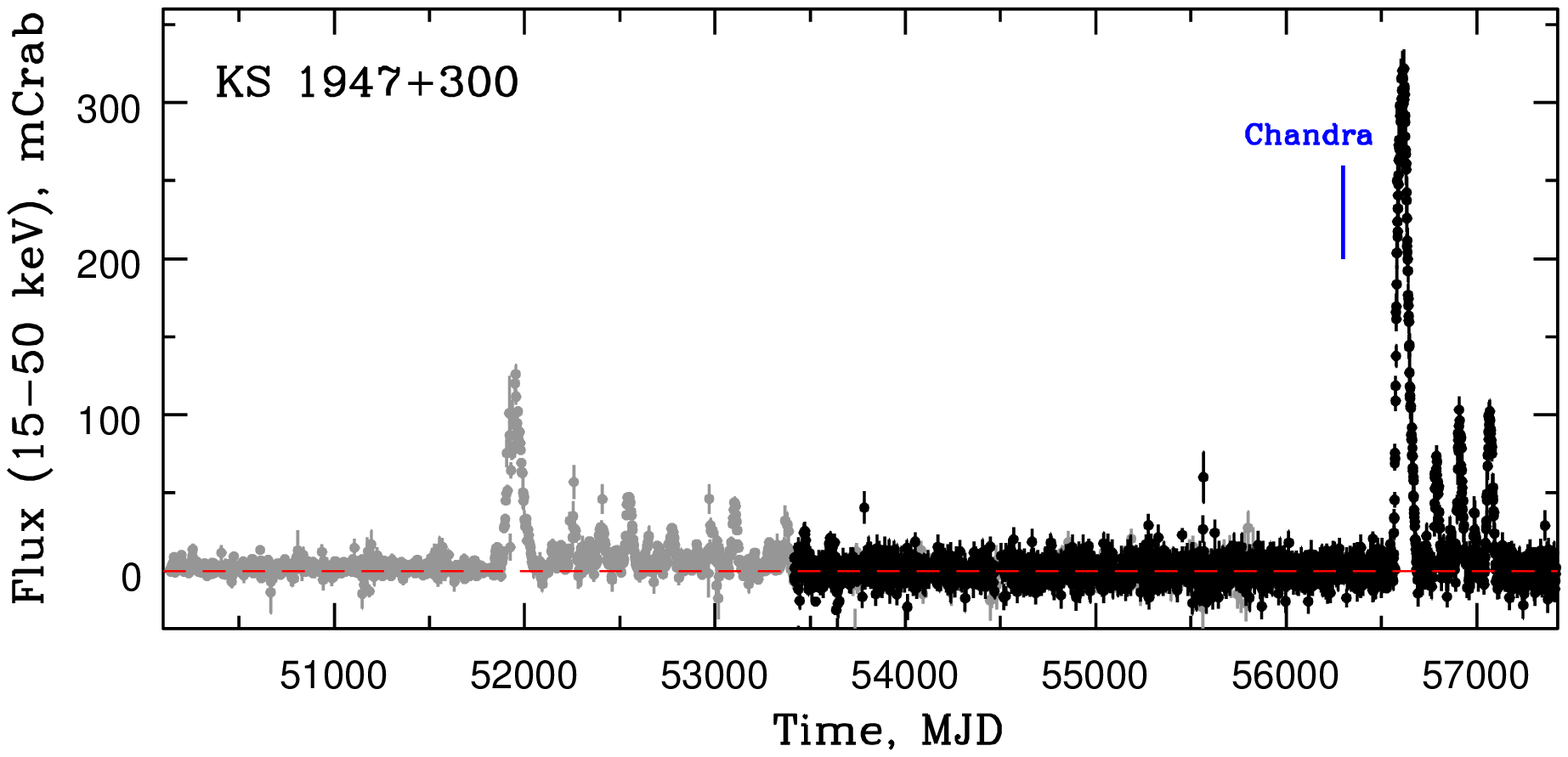}
}
\vspace{3mm}

\hbox{
\includegraphics[height=0.23\textwidth,angle=0,bb=25 415 567 675,clip]{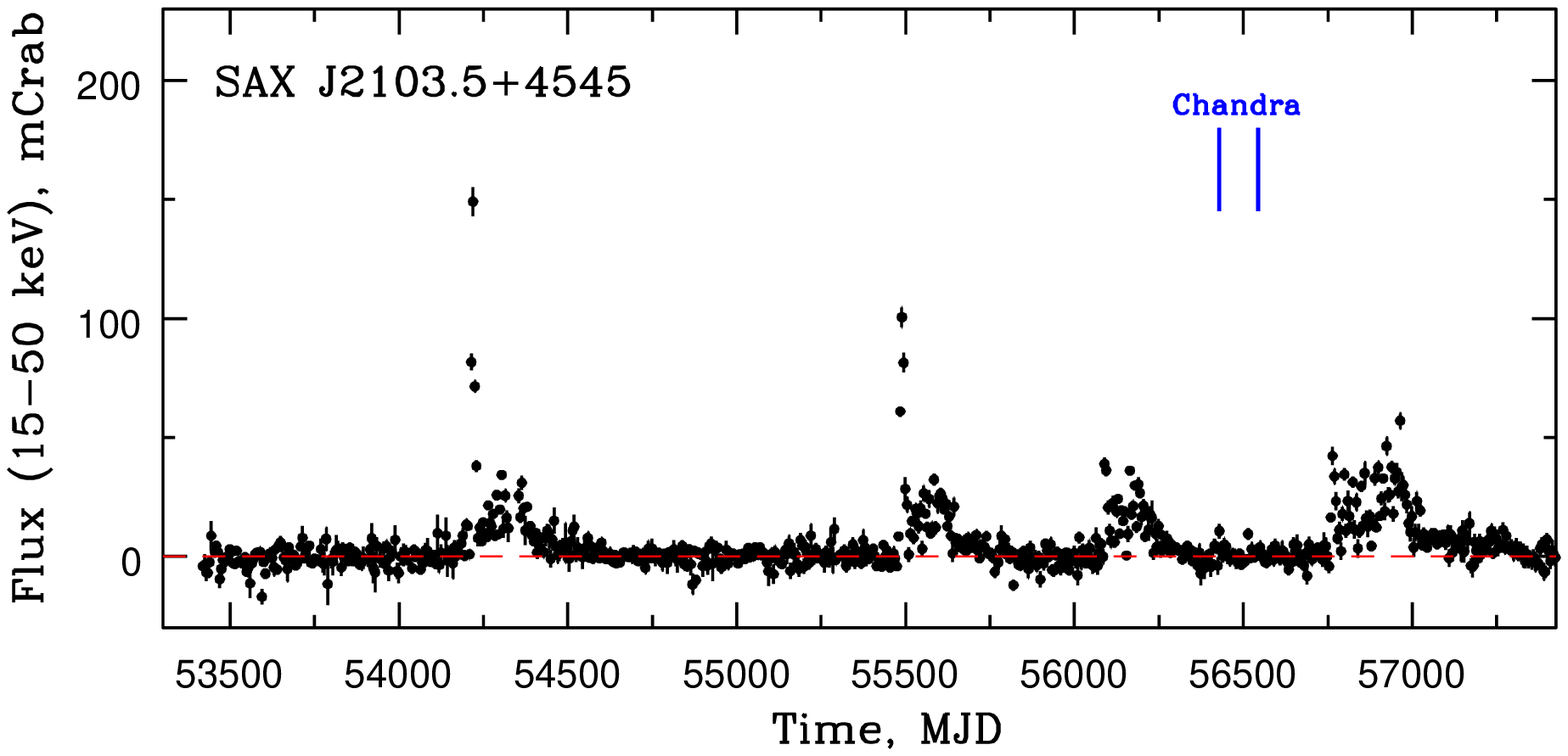}
\hspace{2mm}\includegraphics[height=0.23\textwidth,angle=0,bb=25 415 567 675,clip]{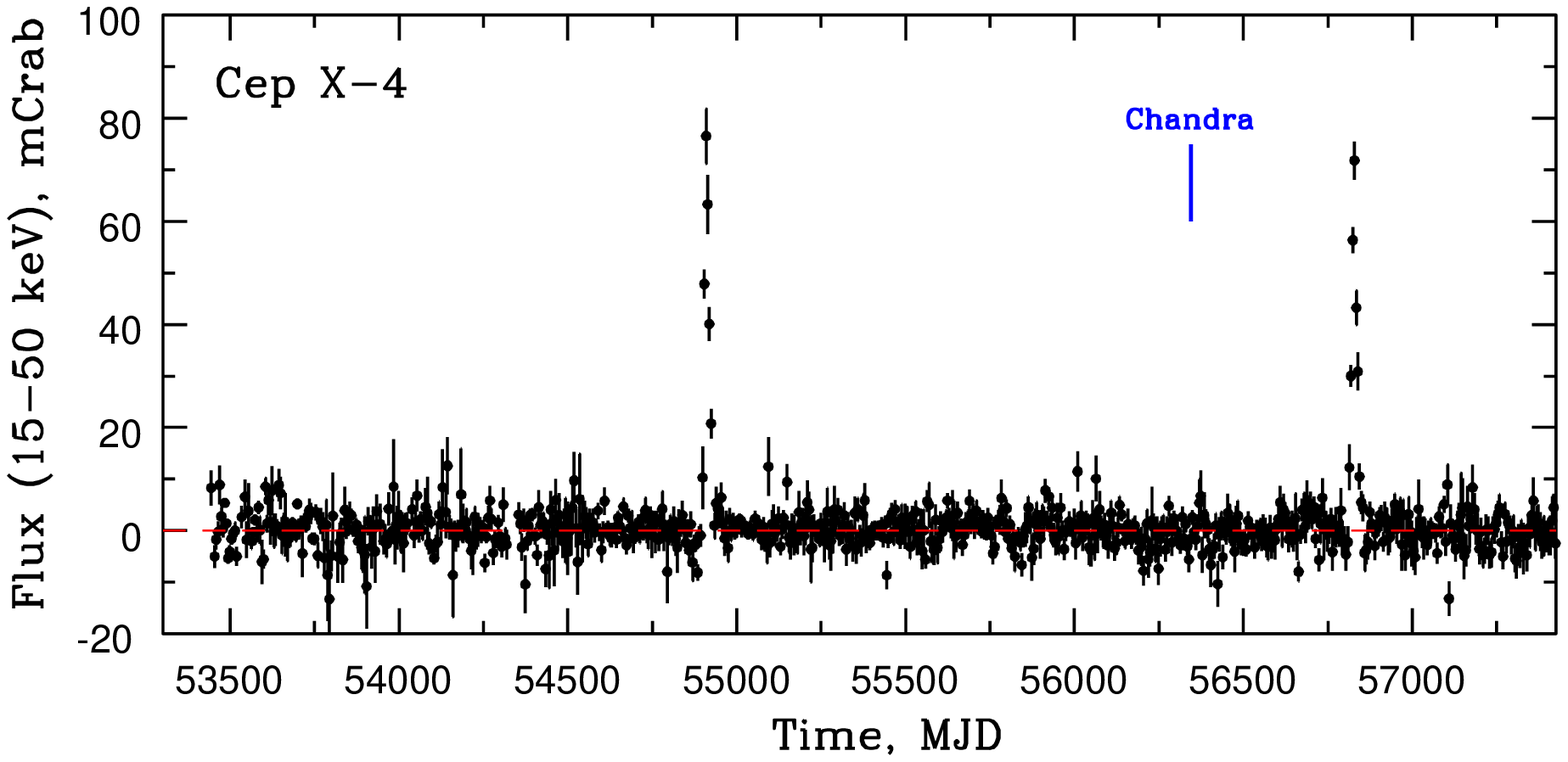}
}
\vspace{3mm}

\contcaption{{\it Swift}/BAT light curves in the 15--50 keV energy
range. The times of the {\it Chandra}, {\it XMM-Newton} and {\it Swift}/XRT
observations utilized in this work are indicated by the vertical bars. {\it
RXTE}/ASM light curves (grey points) in the 2--10 energy range are shown also for several
sources.}
\end{figure*}
%=================================================================

The final spectral and timing analysis for all X-ray instruments was done
using the standard tools of the {\sc ftools/lheasoft 6.17} package. In
particular, the spectral data were fitted in the 0.5--10 keV energy band
using {\sc xspec} package v.12.7.

For fitting the spectra of each source, we used two models: a
power-law model ({\sc powerlaw} in {\sc xspec}, PL) and a blackbody
one ({\sc bbodyrad} in {\sc xspec}, BB). The normalization of the
latter gives the size of the emission region for the known distance to
a source. In both models, absorption by the interstellar medium was
included by using the {\sc phabs} model. Due to the low count rates
(and hence low statistics) for the majority of sources it was
impossible to constrain the absorption columns in our fits. Therefore,
we fixed these at the Galactic interstellar values from
\citet{2005A&A...440..775K}. We note that for several relatively
bright sources we could determine the absorption values with
reasonable accuracy and in all these cases they agreed well with the
Galactic interstellar values. To keep our analysis uniform, we fixed
the absorption columns for these sources to the Galactic values as
well. Finally, taking into account that the vast majority of the
spectra had small number of photons we binned them to have at least 1
count per energy bin and fitted them using W-statistic
\citep{1979ApJ...230..274W}.\footnote{see {\sc xspec} manual;
  \url{https://heasarc.gsfc.nasa.gov/xanadu/}
  \url{xspec/manual/XSappendixStatistics.html}} This is a modification
of the C-statistic \citep{1979ApJ...228..939C} valid if a background
spectrum with Poisson statistics has been read in. In spite of
application of W-statistics, below we use ``$C$-value'' notation
following the common practice.

\subsection{Spectral analysis}
\label{sec:spec}

The results of our spectral analysis are summarized in Table\,\ref{tab:spec_all}.
The table includes: source name; observation ID (ObsID); start time of the
observation (in MJD); effective exposure of the observation; hydrogen column density
$N_{\rm H}$, fixed at interstellar one;
the blackbody temperature $T_{\rm bb}$  and the radius $R_{\rm bb}$ of the emitting area for the {\sc
bbodyrad} model and photon index $\Gamma$ for the {\sc powerlaw} model;
observed flux in the 0.5--10 keV energy band (note, that it is calculated from
the best-fitting model and, therefore is model-dependent); unabsorbed flux in the
0.5--10 keV energy band, calculated using the {\sc cflux} model from the {\sc
xspec} package; unabsorbed luminosity in the same energy band; $C$-statistic
value for both models and corresponding degrees of freedom (dof). The spectra of
our sources are presented in Fig.\,\ref{fig:specs}.

Before proceeding to the discussion of the spectral properties of the
different sources and groups of them, it is necessary to note that for
4U\,0728--25 (ObsID 0003800500N, representing the average of three {\it
Swift}/XRT observations) the start time (Date column) corresponds to the
beginning of the observation 00038005001, and the total exposure is the sum
of the exposures of observations 00038005001, 00038005002, and 00038005003,
which were performed on 2008 July 18, 23 and 25, respectively.
Additionally 4U\,0728--25 was observed with the {\it Chandra} observatory
within the framework of our programme (ObsID 14636). In this observation the
source had an unabsorbed luminosity of about $10^{36}$ \lum\ and a relatively
hard X-ray spectrum with a photon index of $\Gamma\simeq0.5$.  Taking into
account that this paper is dedicated to the study of BeXRPs at very low
luminosities we excluded this bright observation from our subsequent
analysis.

%=================================================================
\begin{figure}
\includegraphics[width=\columnwidth, bb=56 274 550 675,clip]{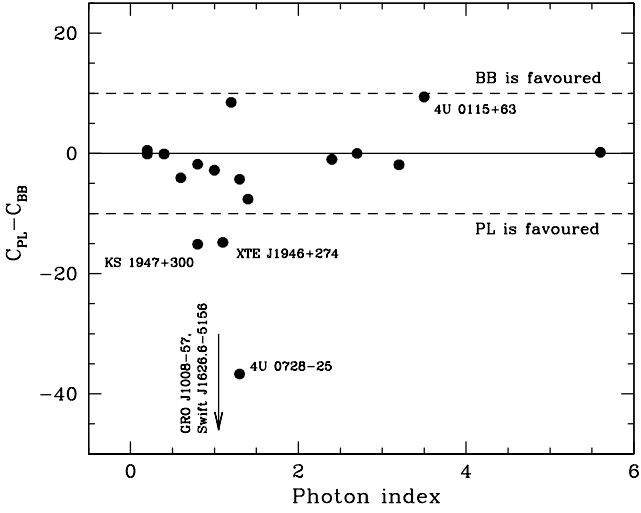}
\caption{Comparison of the fit quality for the PL and BB models for all sources
  from our sample. Two horizontal dashed lines indicate the
  difference $\Delta C=\pm10$, which may be considered as
statistically significant for favouring one of the models over the other.
For two sources GRO\,J1008$-$57 and Swift\,J1626.6$-$5156, $|\Delta C| > 100$,
therefore they are indicated by the arrow.}
\label{fig:cstat}
\end{figure}
%=================================================================

It can be seen from Fig.\,\ref{fig:specs} and Table\,\ref{tab:spec_all} that
for most sources both spectral models provide a more or less satisfactory
fit. To get a quantitative conclusion about the preference of one of the
models for a particular source we used the Akaike information criterion
\citep[AIC; ][]{1974ITAC...19..716A, burnham2011} used both in the cosmological tasks \citep[see
e.g.][]{2007MNRAS.377L..74L} and to discriminate between different spectral
models \citep[see e.g.][]{2017MNRAS.466.1019S}. In our case, AIC$=2k+C$,
where $C$ is the value from the W-statistics, $k$ is number of model parameters
and $k=2$ for both considered
spectral models. 
The difference $\Delta C= C_{\rm PL}-C_{\rm BB} =10$ between
$C$-values yielding from two models implies that one of them is
 $\exp(10/2) \approx 150$ times less probable than the other.

The difference in the $C$-values for the PL and BB models as a function of
the best-fitting power-law index is shown in Fig.\,\ref{fig:cstat}. It is
clearly seen from the figure that $\Delta C=10$ is
reached only for five pulsars where one of the spectral models can be statistically
justified over the other one. However, based on the spectral parameters and AIC value, we can
roughly divide all sources from our sample into several groups. The first
one consists of sources with spectra that are preferentially described by the
power-law model: 4U\,0728$-$25, GRO\,J1008$-$57, Swift\,J1626.6$-$5156,
XTE\,J1946+274, KS\,1947+300, {\it and possibly } GS\,0834$-$43, Cep\,X-4
and SAX\,J2239.3+6116. This group harbours the brightest objects from our
sample with a luminosity around $10^{34}$\,\lum, as well as several fainter systems
(down to a few times $10^{33}$\,\lum). The typical value of the photon
index for these sources is around $\Gamma\sim1$. The second group includes
sources with thermal spectra: 4U\,0115+63, V\,0332+53 ( for both sources 
the $\Delta C$ value is just slightly below 10), and
probably RX\,J0812.4$-$3114,  MXB\,0656$-$072.  The temperature of the emission in these systems
(using the blackbody model) is $kT\simeq0.1-0.4$ keV, and the luminosity is
very low, a few $10^{32}-10^{33}$\,\lum.  Note that radii of the
emission regions for 4U\,0115+63 and V\,0332+53 are several hundred metres,
which agree well with expected sizes of polar caps of the NS
\citep[see e.g.][]{2015MNRAS.447.1847M}. Due to very limited statistics, we
could not restrict the size of the emission region for RX\,J0812.4$-$3114,
which has the softest spectrum. A third group includes two sources with a
moderate luminosity $(2-5)\times10^{33}$\,\lum, for which it is difficult to
unambiguously determine their spectral shape:
 2S\,1417$-$624 and SAX\,J2103.5+4545.

We note that SAX\,J2103.5+4545 was observed with the {\it Chandra}
observatory in 2013 Sep (MJD 56544) with a very long exposure ($\sim45$ ks)
in a slightly lower intensity state in comparison to our observations. The
source spectrum was well described with an absorbed blackbody model with a
temperature of $kT\simeq1$ keV and radius of $R=0.11$ km
\citep{2014MNRAS.445.1314R}, which agrees well with the results of our
measurements (Table\,\ref{tab:spec_all}). Therefore, it is likely that also
during our observation this source had indeed a spectrum that could be best
described by the blackbody model. The count rates for the last two sources
-- 2S\,1553$-$542 and GS\,1843+00 -- are extremely low and prevents us from
performing detailed spectral analysis inhibiting us from putting them in any of the
abovementioned groups.

\subsection{Timing analysis}
\label{sec:time}

Along with the spectral analysis we also performed a timing analysis to
search for pulsations from all sources in our sample. As a first step all
light curves were corrected to the Solar system barycentre.  The corrected
light curves in the 0.5--10 keV energy range were investigated with the {\sc
efsearch} tool for the presence of a coherent signal. As a result, we
detected pulsations from five sources: Swift\,J1626.6$-$5156, XTE\,J1946+274,
KS\,1946+300, SAX\,J2103.5+4545 and Cep\,X-4. For four of them (except
SAX\,J2103.5+4545) pulsations at such low luminosities are detected for the
first time. Corresponding pulse profiles are shown in Fig.\,\ref{fig:pprof}
(zero phase is arbitrary).  It is interesting to note that all sources show
quite a high pulsed fraction\footnote{It was determined as
$\mathrm{PF}=(F_\mathrm{max}-F_\mathrm{min})/(F_\mathrm{max}+F_\mathrm{min})$,
where $F_\mathrm{max}$ and $F_\mathrm{min}$ are the maximum and minimum
flux in the pulse profile, respectively.}  -- between 50 and 70 per cent with
a characteristic uncertainty of $\sim$15 per cent. Such high pulsed fraction values
are typical for X-ray pulsars both at low accretion rate \citep[see
e.g.][]{2009AstL...35..433L} and in the quiescent state \citep[see
e.g.][]{2000A&A...356.1003N,2014MNRAS.445.1314R}. Therefore, it is difficult
to make any final conclusions about the origin of the emission in the low
state based only on the pulsed fraction value. Very low count statistics and small
exposure times did not allow us to put meaningful upper limits on the
pulsations presence in the other sources.

%=================================================================
\begin{figure*}
\centering

\hbox{
\includegraphics[height=0.193\textwidth,angle=0,bb=20 305 545 695,clip]{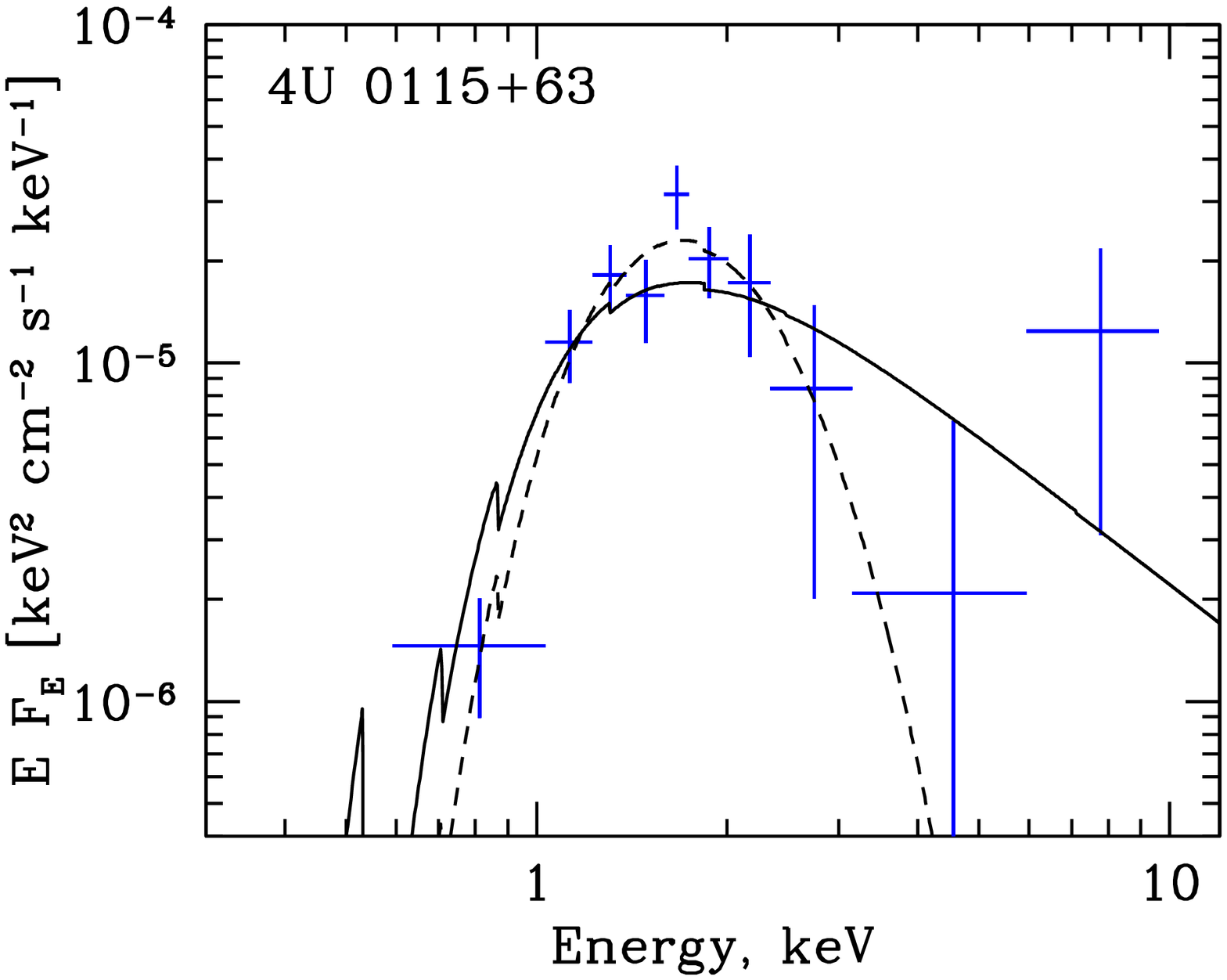}
\hspace{2mm}\includegraphics[height=0.19\textwidth,angle=0,bb=48 305 545 690,clip]{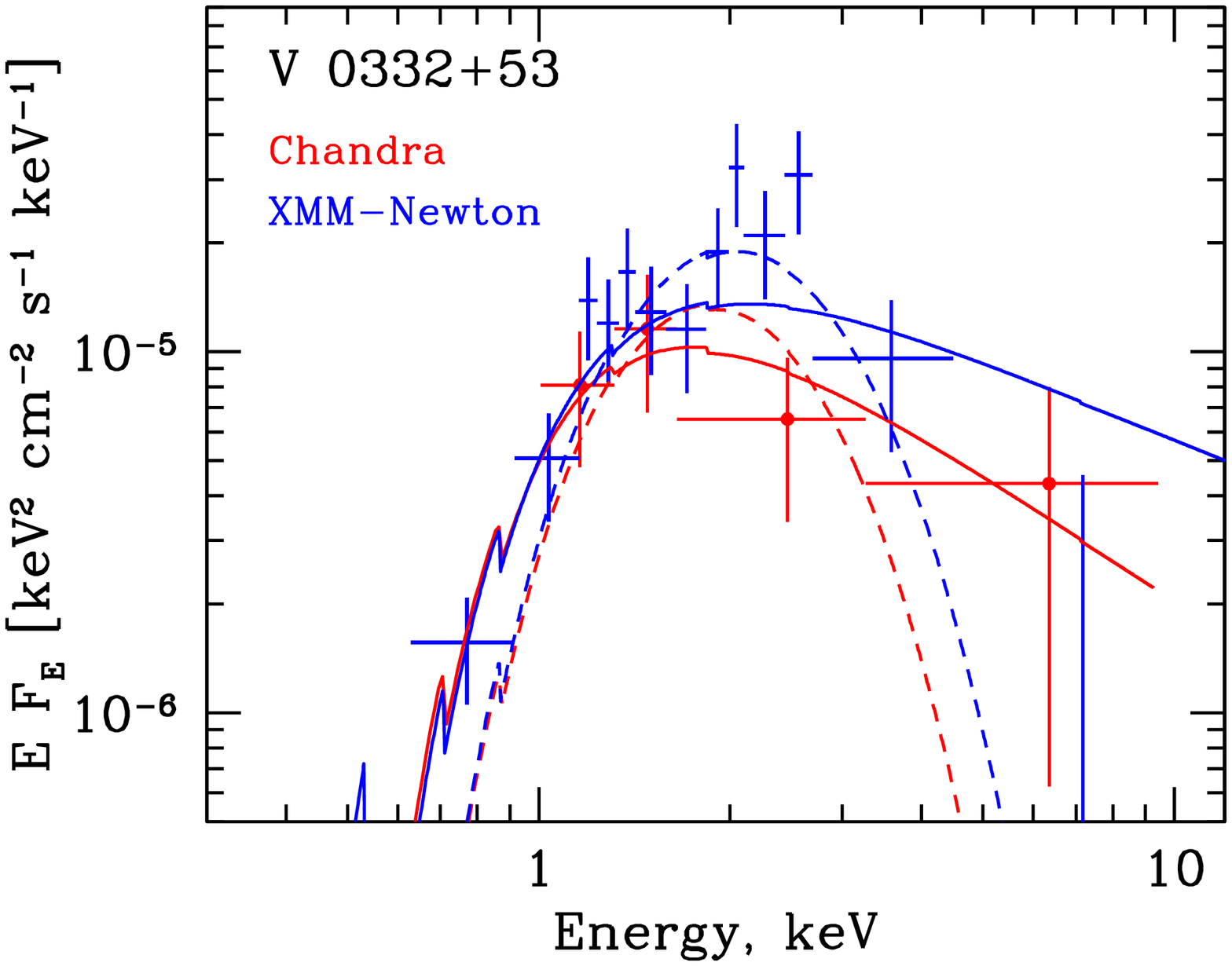}
\hspace{2mm}\includegraphics[height=0.19\textwidth,angle=0,bb=48 305 545 690,clip]{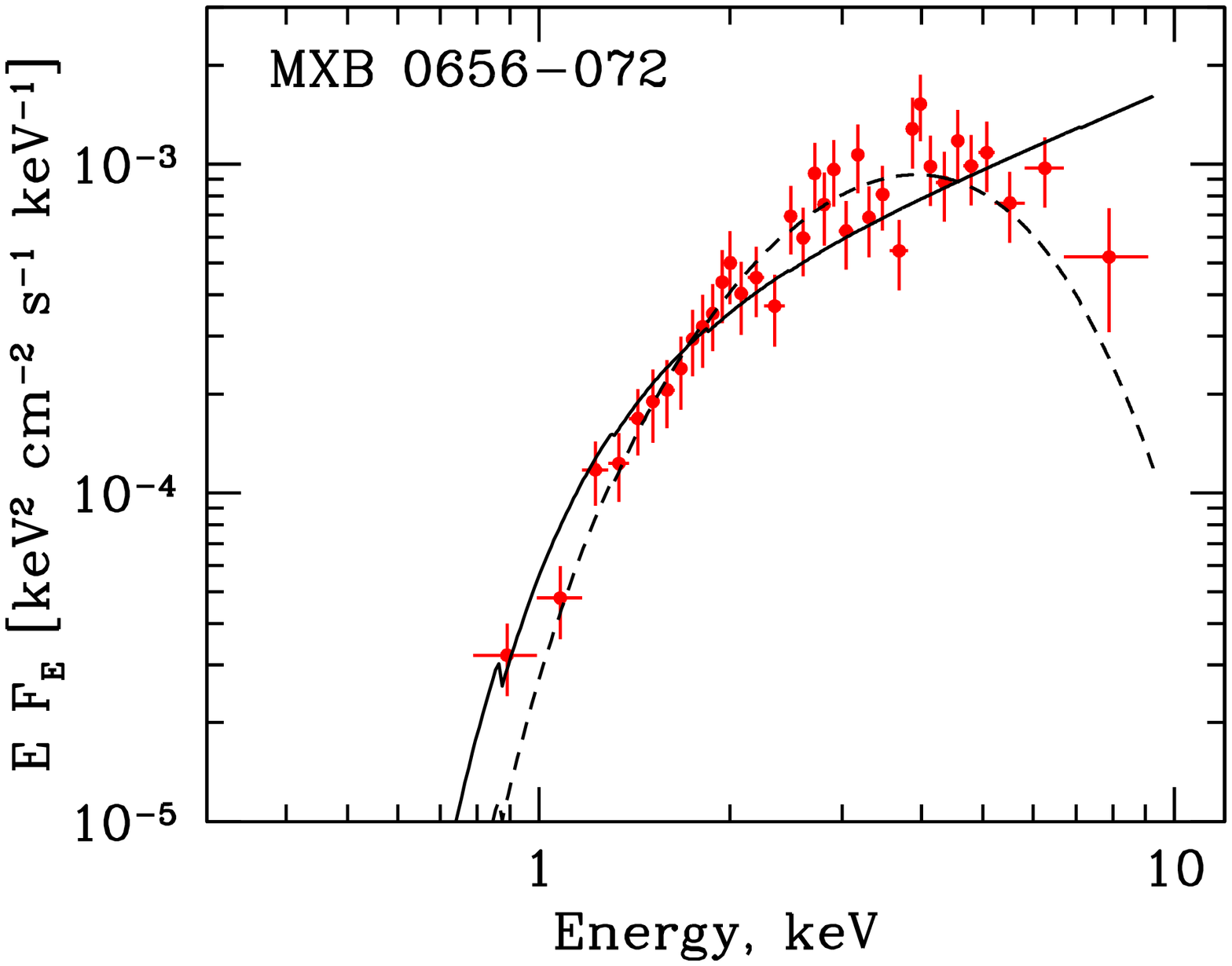}
\hspace{2mm}\includegraphics[height=0.19\textwidth,angle=0,bb=48 305 545 690,clip]{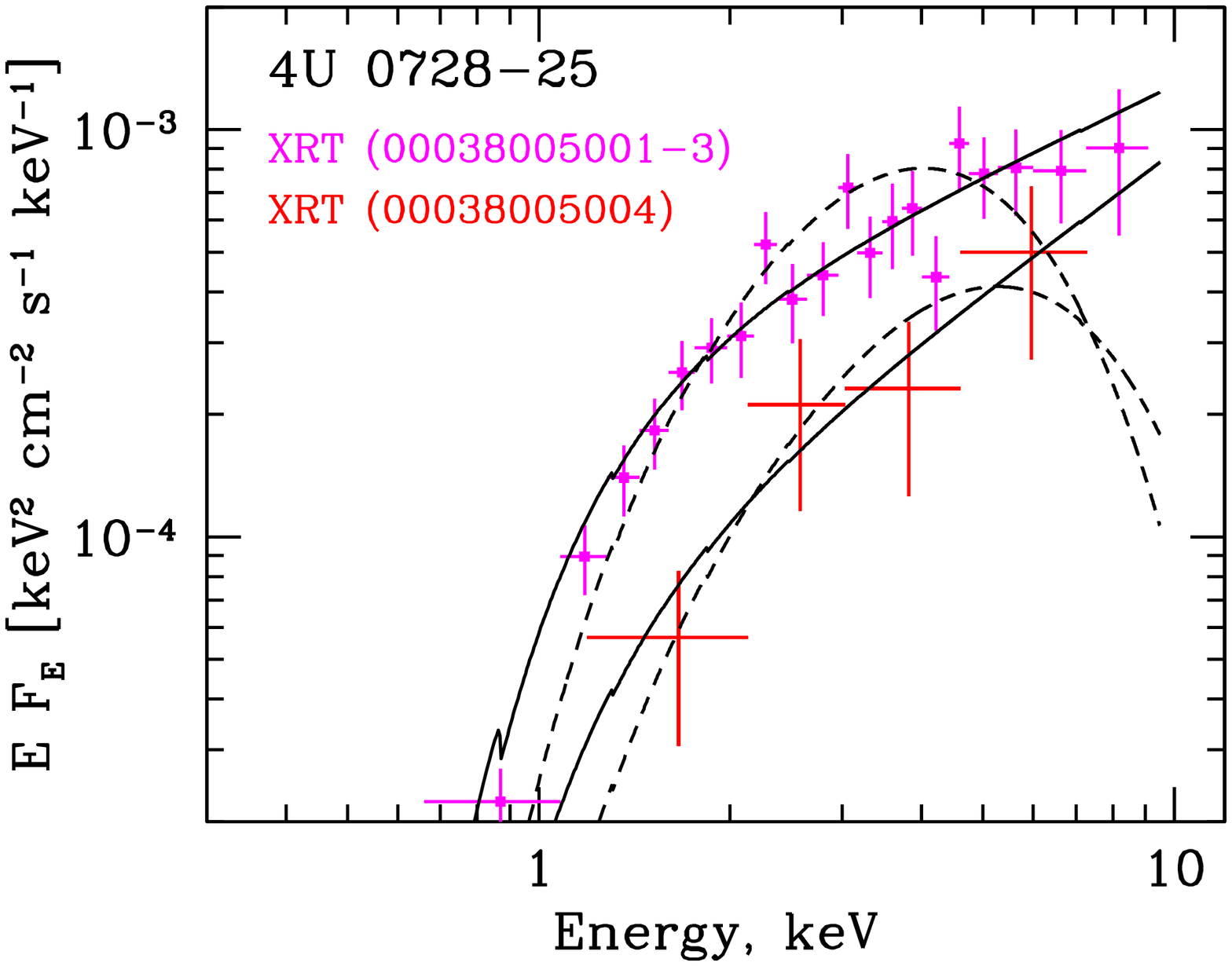}
}
\vspace{3mm}

\hbox{
\includegraphics[height=0.19\textwidth,angle=0,bb=20 305 545 690,clip]{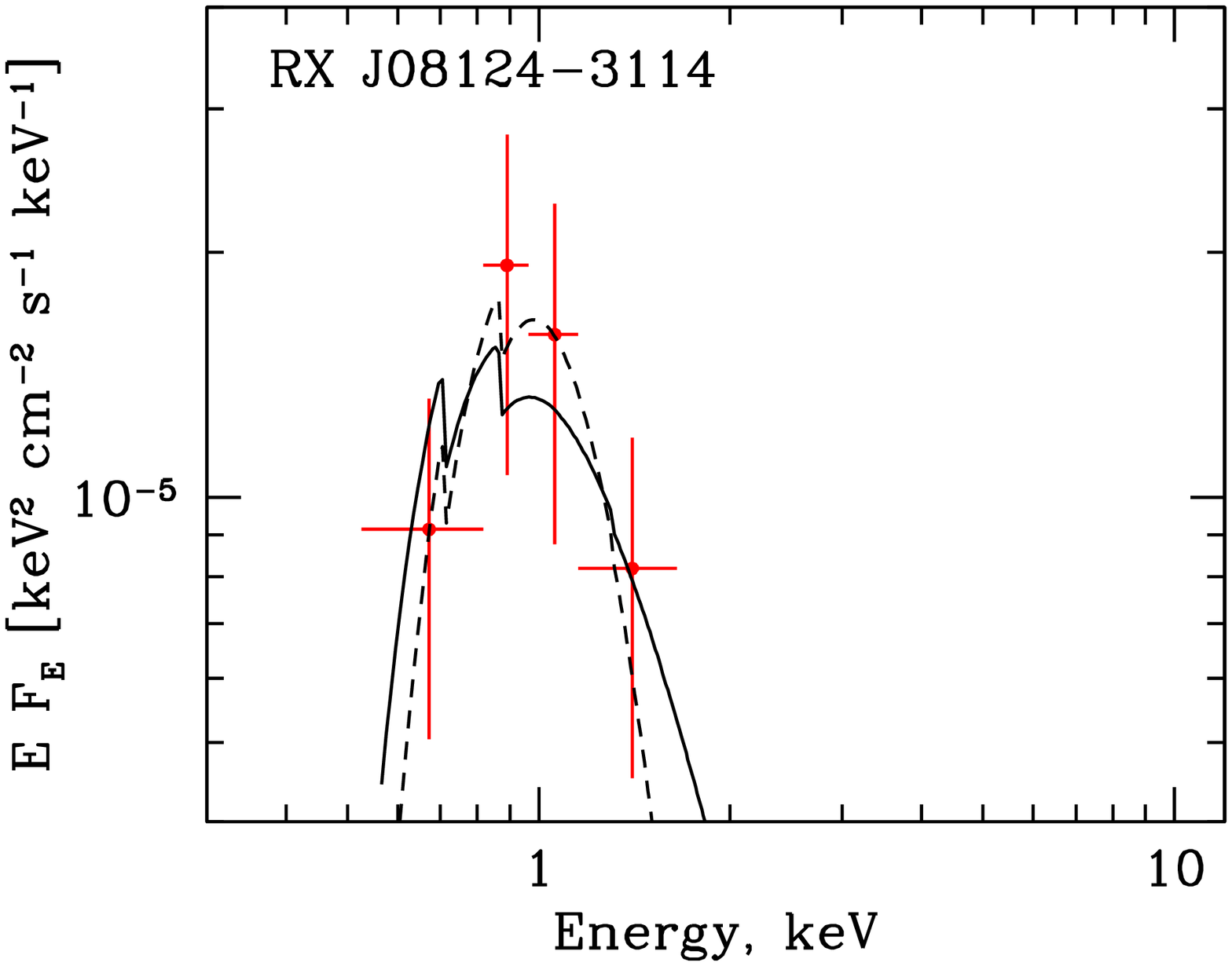}
\hspace{2mm}\includegraphics[height=0.19\textwidth,angle=0,bb=48 305 545 690,clip]{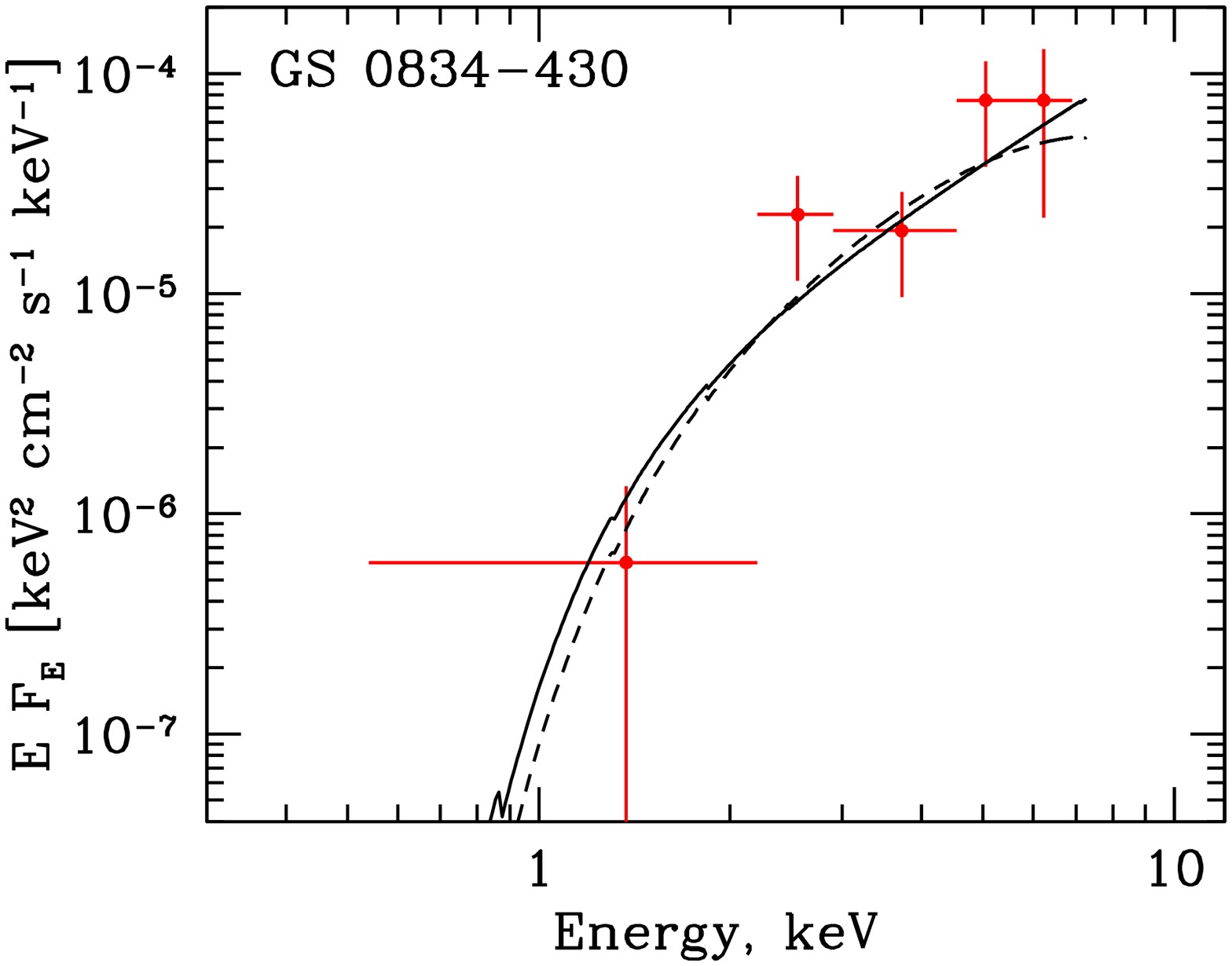}
\hspace{2mm}\includegraphics[height=0.19\textwidth,angle=0,bb=48 305 545 690,clip]{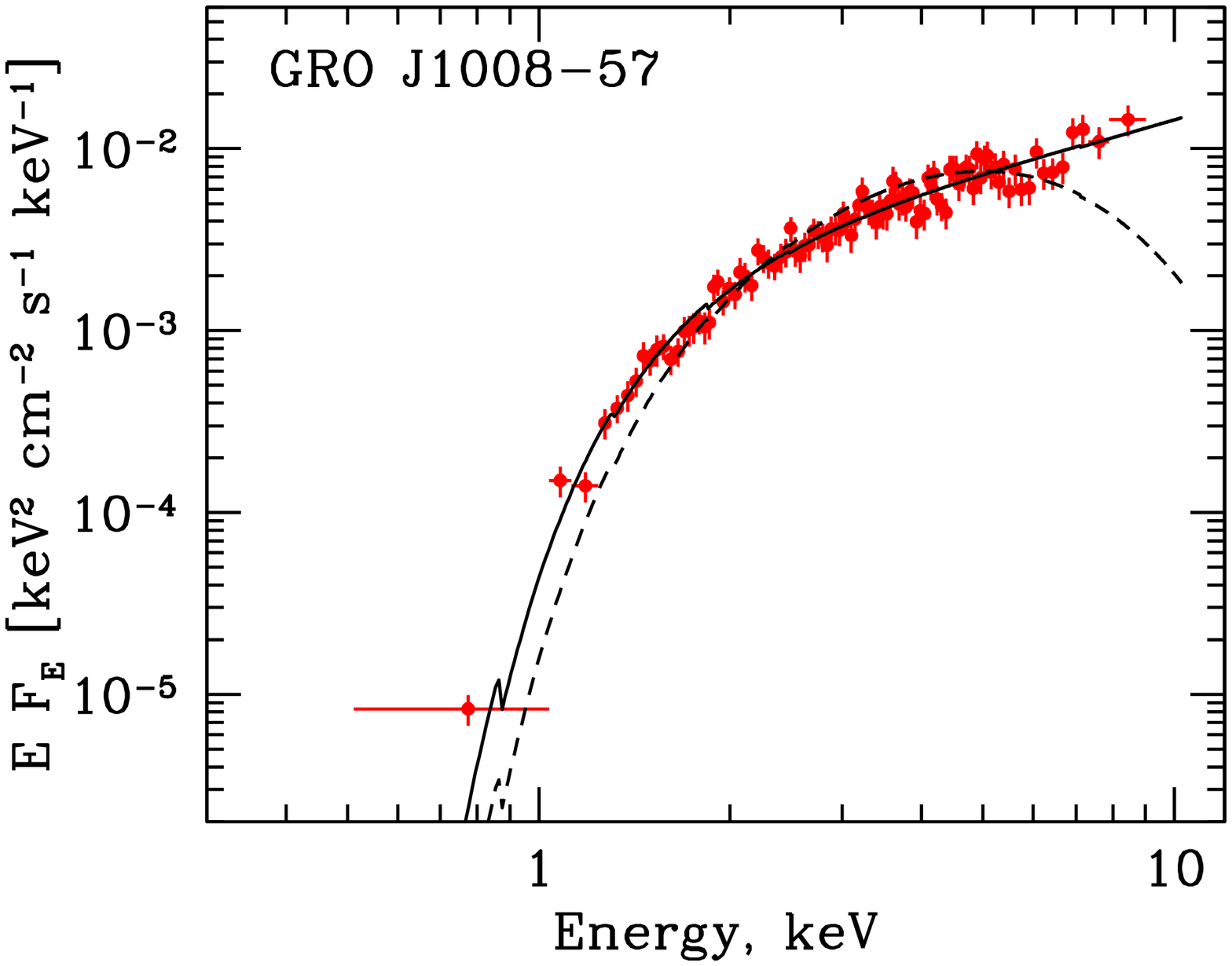}
\hspace{2mm}\includegraphics[height=0.19\textwidth,angle=0,bb=48 305 545 690,clip]{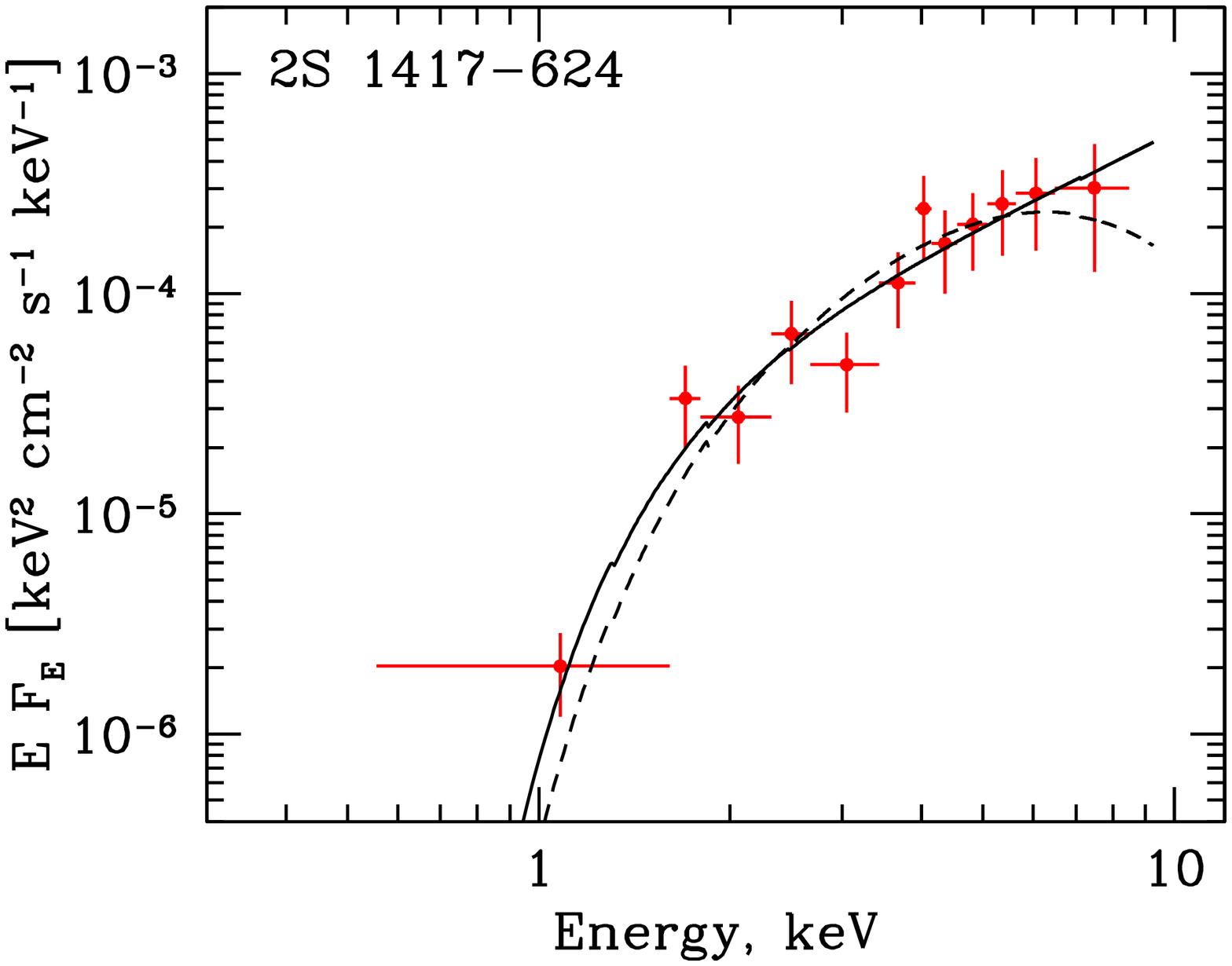}
}
\vspace{3mm}

\hbox{
\includegraphics[height=0.19\textwidth,angle=0,bb=20 305 545 690,clip]{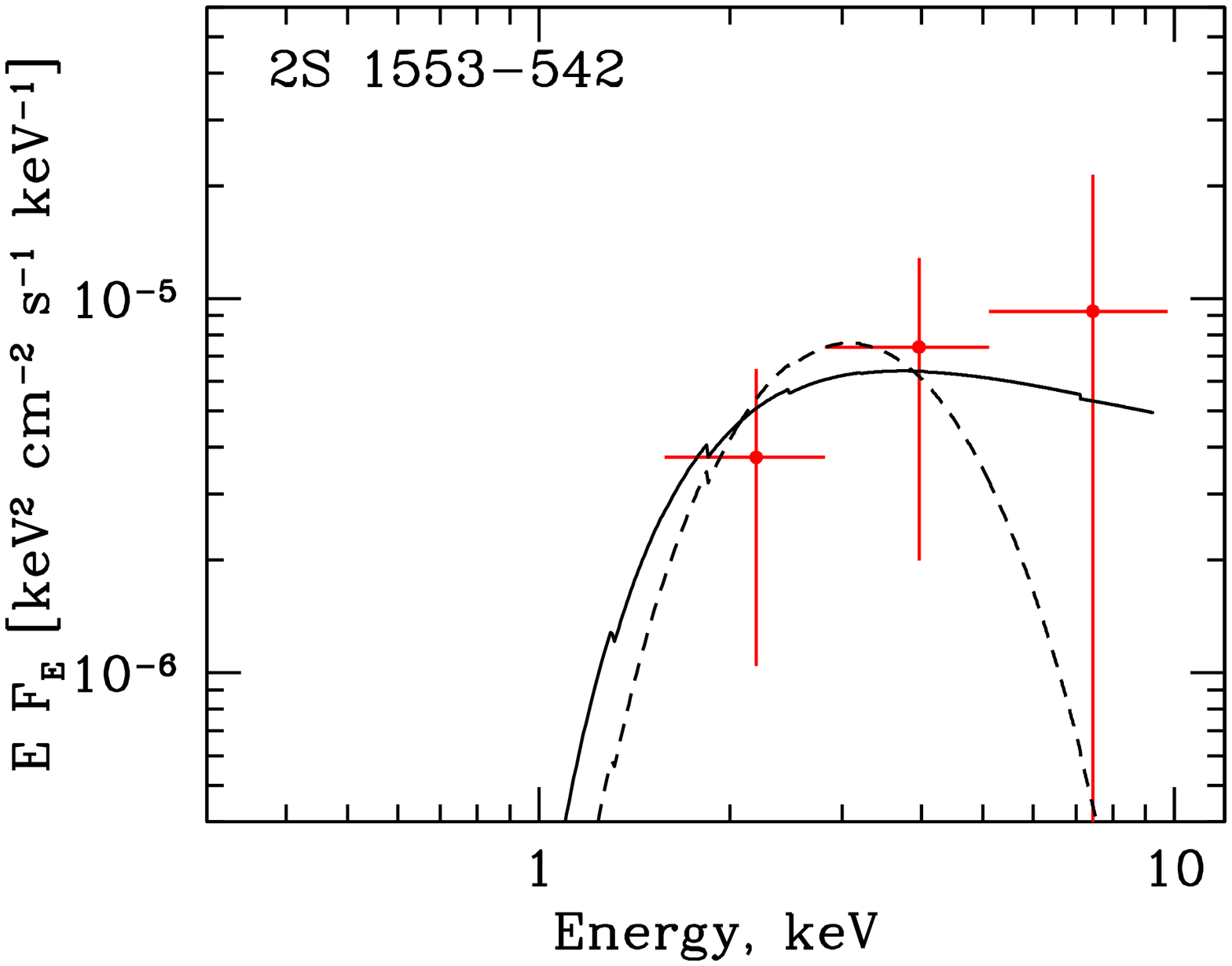}
\hspace{2mm}\includegraphics[height=0.19\textwidth,angle=0,bb=48 305 545 690,clip]{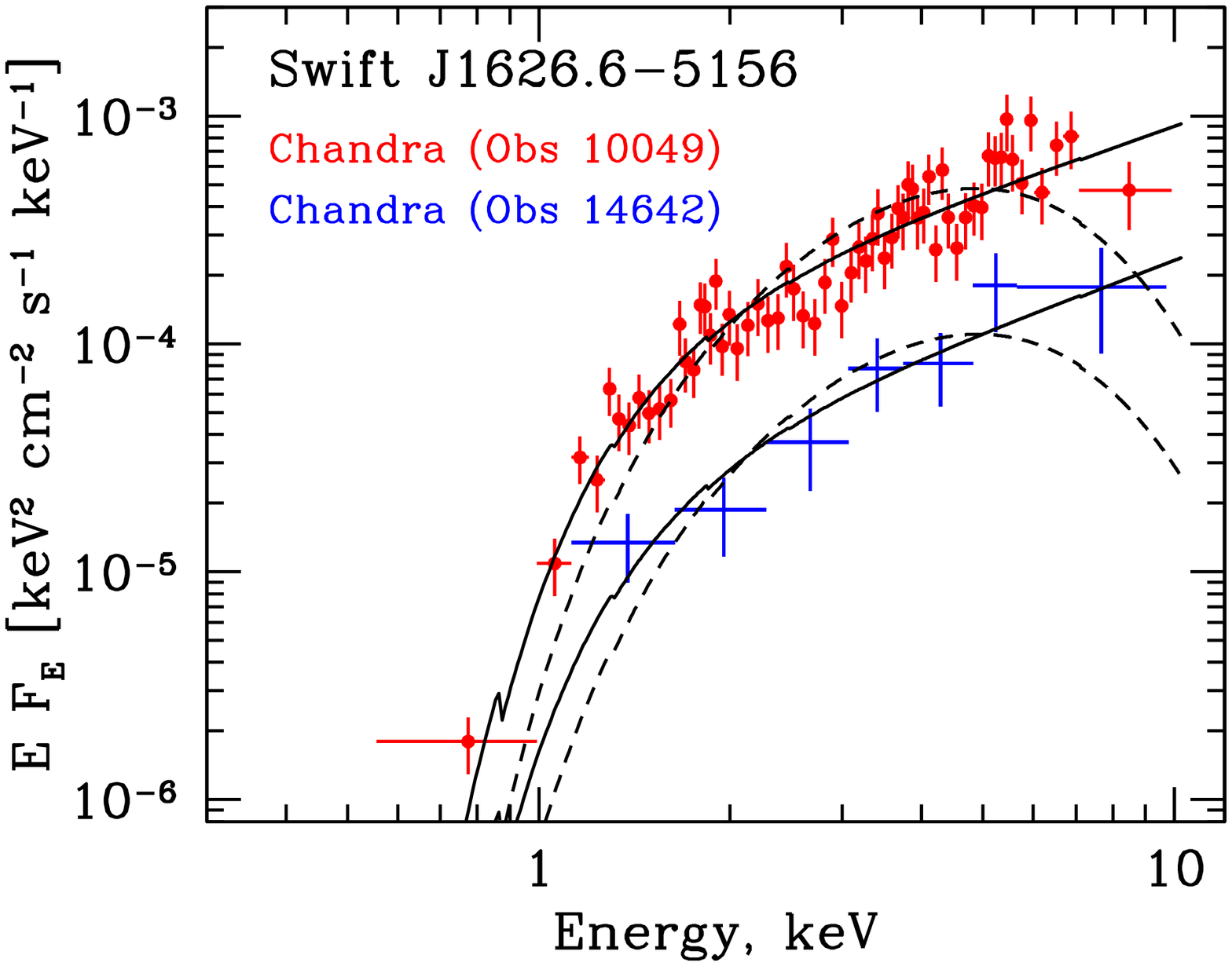}
\hspace{2mm}\includegraphics[height=0.19\textwidth,angle=0,bb=48 305 545 690,clip]{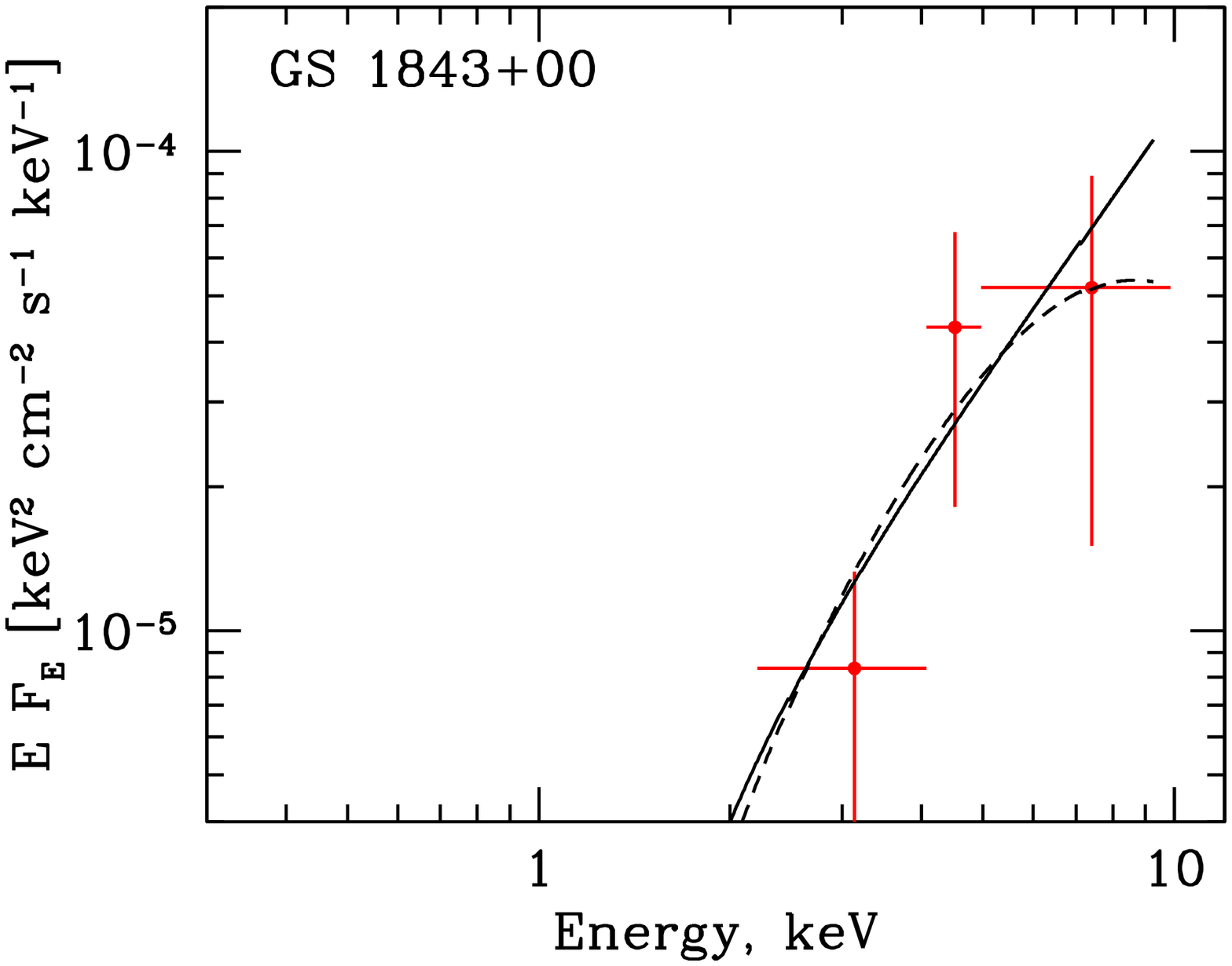}
\hspace{2mm}\includegraphics[height=0.19\textwidth,angle=0,bb=48 305 545 690,clip]{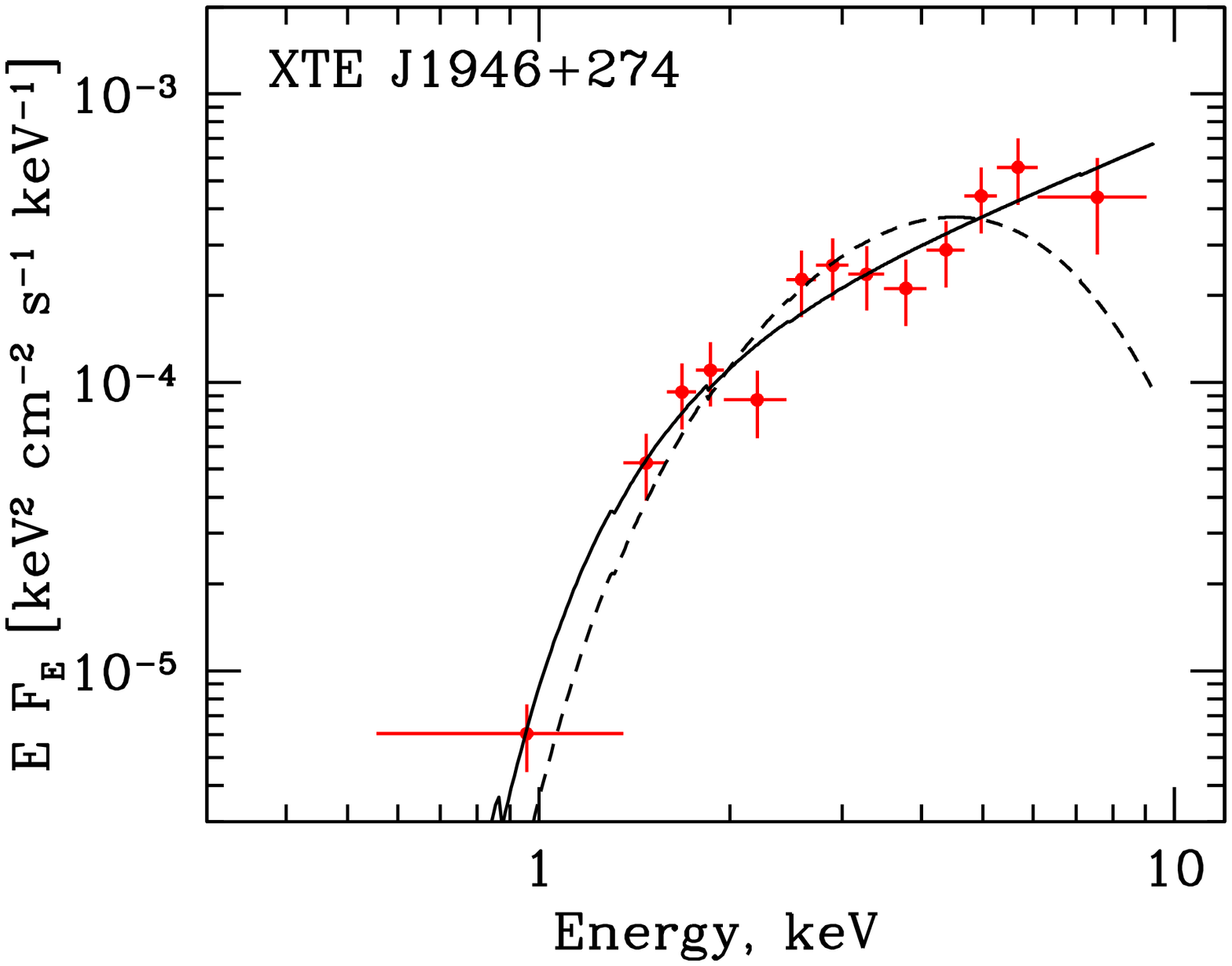}
}
\vspace{3mm}

\hbox{
\includegraphics[height=0.204\textwidth,angle=0,bb=20 275 545 690,clip]{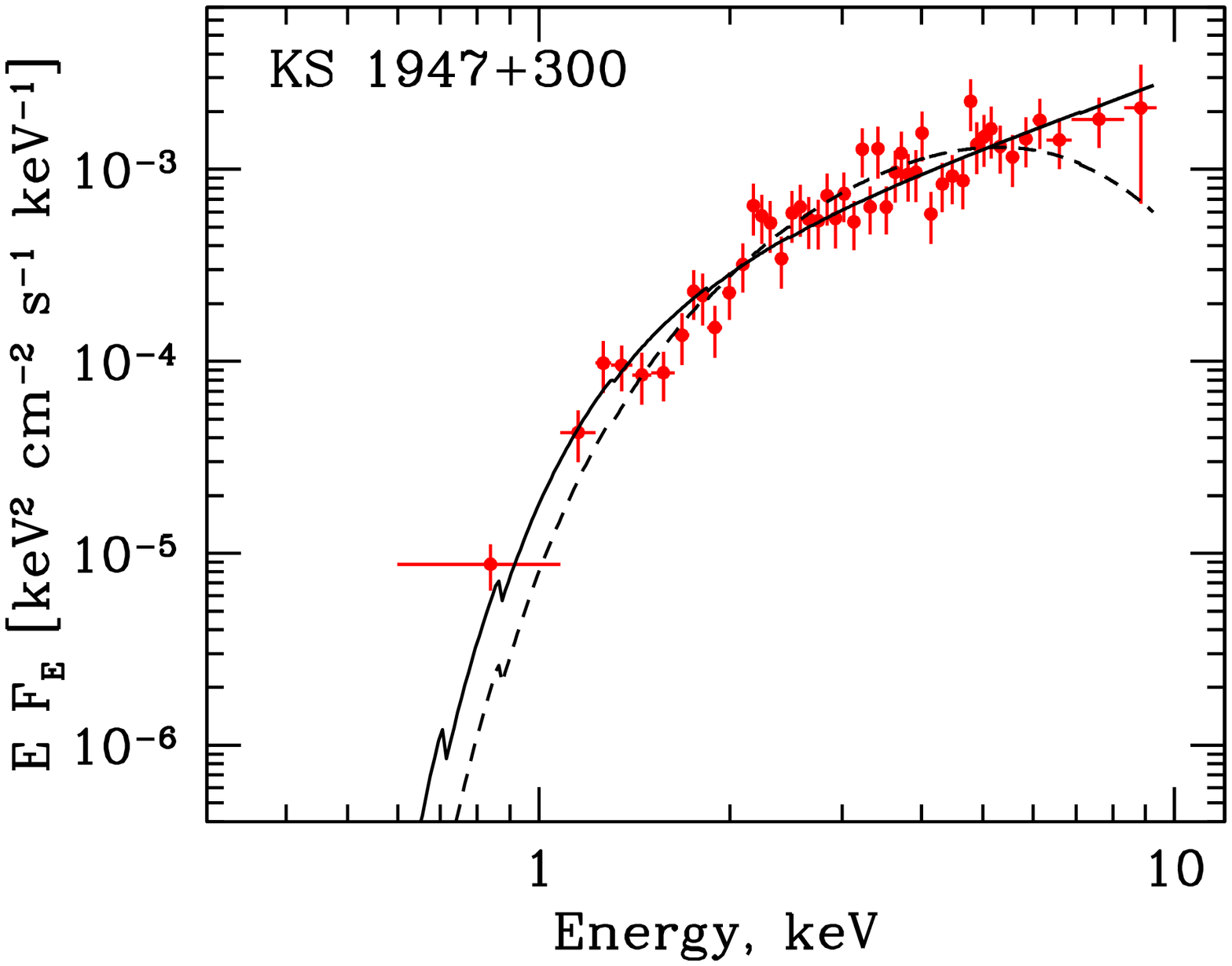}
\hspace{2mm}\includegraphics[height=0.204\textwidth,angle=0,bb=48 275 545 690,clip]{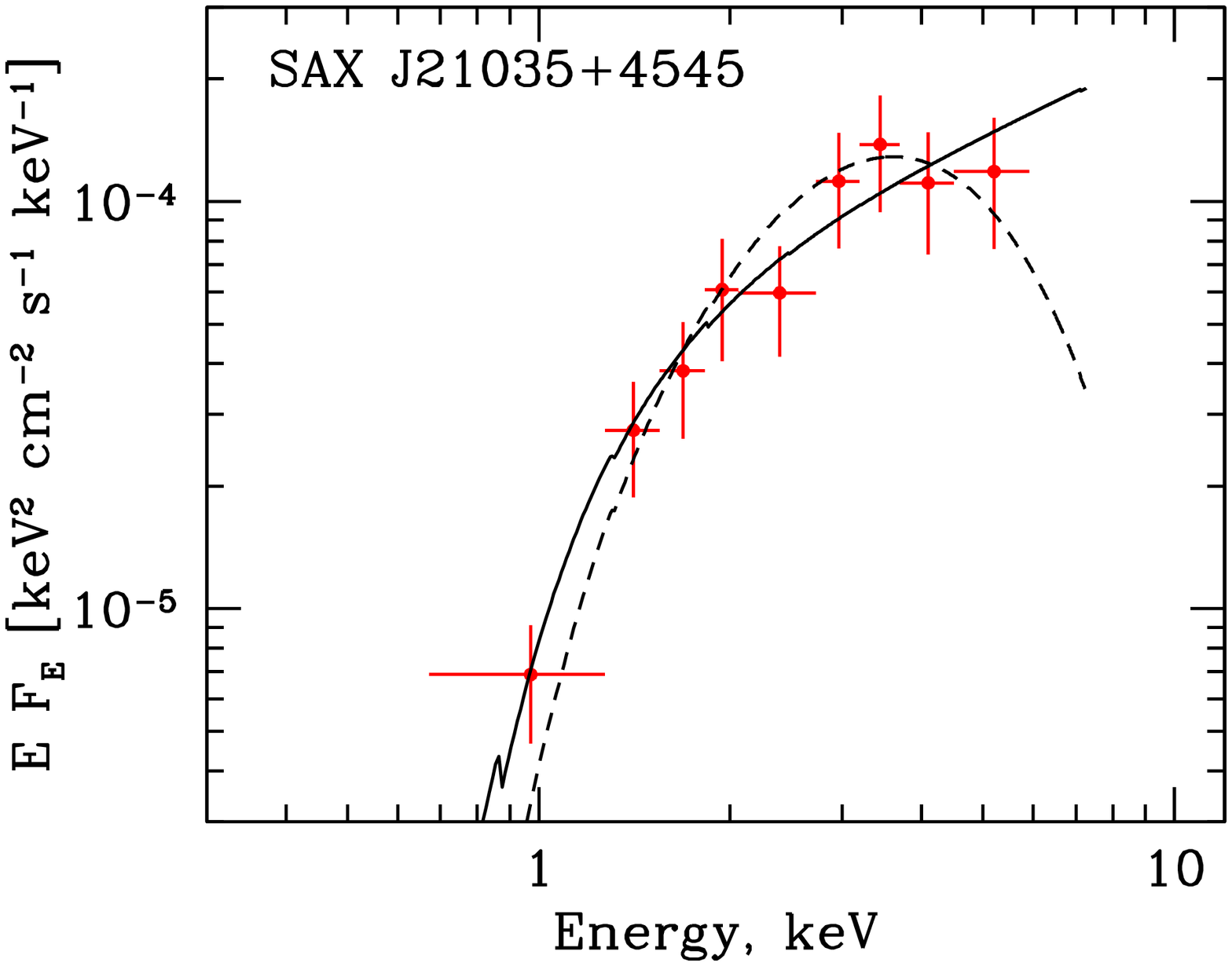}
\hspace{2mm}\includegraphics[height=0.204\textwidth,angle=0,bb=48 275 545 690,clip]{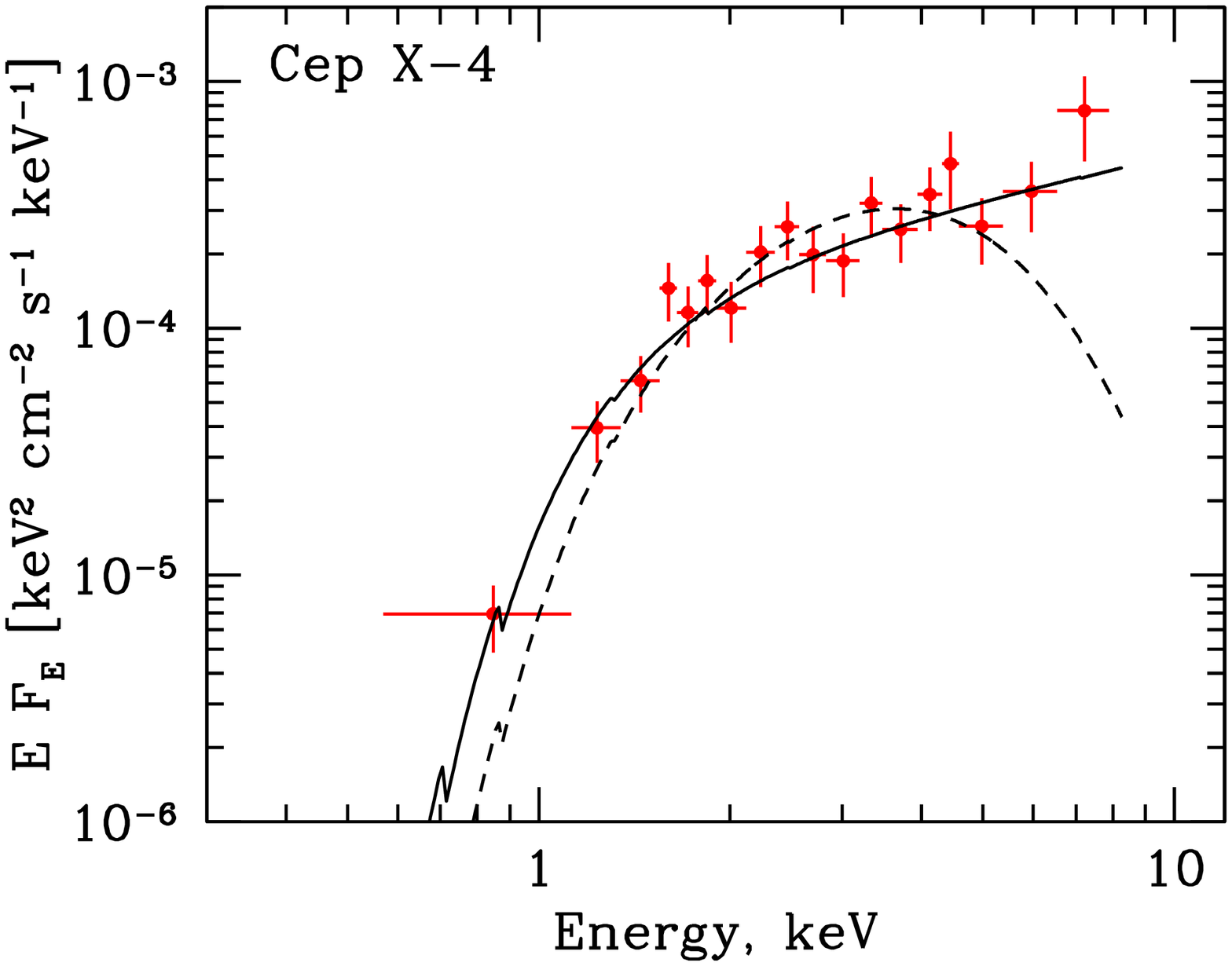}
\hspace{2mm}\includegraphics[height=0.204\textwidth,angle=0,bb=48 275 545 690,clip]{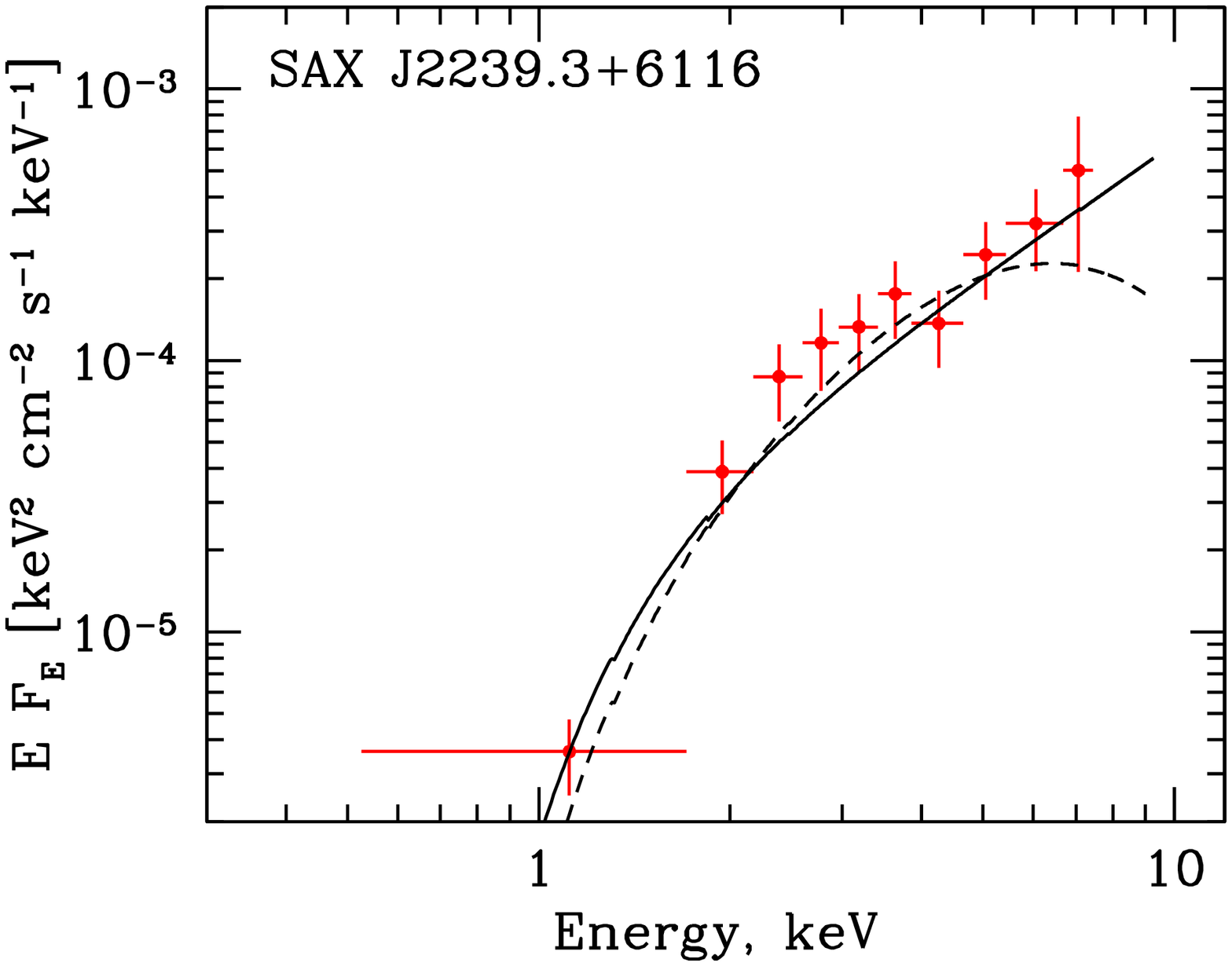}
}

\caption{Unfolded X-ray spectra for all the sources in our sample.
Solid and dashed lines represent the power-law and blackbody best fitting models (both
modified by absorption), respectively. Data obtained with {\it Chandra},
{\it XMM-Newton} and {\it Swift}/XRT are shown by red circles, blue crosses
and magenta squares, respectively, unless stated otherwise.}\label{fig:specs}
\end{figure*}
%=================================================================

\section{Individual sources}
\label{indiv}

In this section we describe the observational properties of each source and
compare them with predictions of the relevant models presented in
Section~\ref{discussion} (the corresponding predictions for the propeller
limiting luminosity, averaged mass accretion rate and expected quiescent
luminosity are introduced in Section~\ref{discussion} and listed in Table
\ref{tablum}). References for the results of previous observations in the
quiescent state are presented as well (if available). \\

\subsection{4U\,0115+63}

{\it XMM-Newton} observed 4U\,0115+63 about three years after a
powerful type II outburst that occurred in 2004 September--October.
The source has a thermal spectrum (see Fig. \ref{fig:specs}) with the
temperature of about 0.3 keV and a luminosity around $6\times10^{32}$
\lum. This luminosity is a factor of $\sim10$ lower than was observed
by the {\it Swift}/XRT telescope immediately after another giant
outburst in 2015 and the subsequent transition of the source to the
propeller regime with the typical temperature of 0.5--0.9 keV
\citep{2016A&A...593A..16T,2016MNRAS.463L..46W}. Our quiescent
luminosity is consistent (within the uncertainties) with the one found
by \cite{2002ApJ...580..389C} during a BeppoSAX observation of the source.

Due to a long period of a quiescence before the {\it XMM-Newton} observation
it is quite possible that the crust of the NS was in thermal equilibrium with
the core \citep{2002ApJ...580..413R}. However, as seen from Fig.
\ref{fig:lcs}, the type II outbursts from this source are observed every few
years and it is not clear if equilibrium can be achieved on this times-cale.
Nevertheless, 4U\,0115+63 was observed by {\it XMM-Newton} with the lowest
value of the blackbody temperature.

\subsection{V\,0332+53}

Both {\it Chandra} and {\it XMM-Newton} observed V\,0332+53 in the very
low state. The {\it Chandra} observation was performed at least
five years after any strong outburst activity. However, the period of
quiescence could have been much longer, as the {\it RXTE}/ASM monitor started
to operate only in 1996 (before that a bright outburst was observed in 1989).
The source spectrum is well fitted with the blackbody
model with a temperature of $\sim0.4$ keV and a luminosity
$\sim3\times10^{32}$ \lum. Similar to 4U\,0115+63 this luminosity is
a factor of $\sim10$ lower than observed from the heated surface of
the NS right after its transition to the propeller regime
\citep{2016A&A...593A..16T,2016MNRAS.463L..46W}.

Note that during the {\it XMM-Newton} observation performed about three years
after the type II outburst in 2004 Dec to 2005 Jan, the source luminosity
and temperature of its blackbody emission were roughly the same as during
the {\it Chandra} observation. Spectral parameters derived in our analysis
are consistent within the uncertainties  with the results of an independent
analysis by \citet{2016MNRAS.463...78E} and previous results of
\citet{2002ApJ...580..389C}.

Our measurements can be used to estimate roughly the characteristic time
for the crust to reach thermal equilibrium again with the core (assuming that
indeed the accretion of matter during the preceding type-II outburst lifted
the crust out of equilibrium). Because the luminosities and temperatures of the
source during the {\it Chandra} and {\it XMM-Newton} observations were
consistent with being the same, we can assume that indeed the NS
crust was in equilibrium with the core (else, it is quite likely that after
the {\it XMM-Newton} observation the source would have displayed a hotter
crust). Note, that likely the last outburst preceding the {\it Chandra}
observation was detected in 1989, i.e. $\sim15$ years before. At the same
time the {\it XMM-Newton} observation was performed about 3 years after the
preceding type-II outburst so this would then be an upper limit on the
characteristic equilibration time.

\subsection{MXB\,0656$-$072}

The {\it Chandra} observation was performed about four years after a sequence of
outbursts that started in 2008 October. Based on the above
arguments outlined for V\,0332+53 on the equilibration time-scale,
 in  MXB\,0656$-$072 the crust may have been in equilibrium
with the core at the time of our {\it Chandra} observation (note this assumes
that the equilibration time-scale does not vary significantly between the sources).
This is supported by the spectral analysis revealing a thermal shape of the
spectrum with the blackbody temperature about 1 keV and the luminosity
$3.8\times10^{33}$ \lum. At the same time it is about 10 times higher in
comparison to the limiting luminosity for the propeller regime onset (see Section~\ref{prop} and
Table \ref{tablum}). This discrepancy can be understood in the frame of the cold
accretion disc paradigm \citep[see Section~\ref{sec:cold}; ][]{tsyg2017}. Indeed, assuming the same luminosity decline law as obtained in the quoted paper $L \propto t^{-0.7}$, a transition to the cold accretion regime around $10^{35}$ \lum, and absence of transient activity after 2008 outburst, one can expect a luminosity of MXB\,0656$-$072 around a few times $10^{32}$ \lum\ at the moment of our {\it Chandra} observation, which is below the observed one.
In this case the quiescent luminosity estimated from the deep-crustal heating model ($\sim1.4\times10^{33}$~\lum\ for this pulsar; see Section~\ref{sec:heating}) should be the dominant source of emission and naturally explains the soft shape of the energy spectrum.

\subsection{4U\,0728$-$25}

This source did not exhibit strong outbursts neither in the {\it RXTE}/ASM
nor the {\it Swift}/BAT data. However, both {\it Swift}/XRT and {\it Chandra}
observations reveal a bright source with a hard spectrum that is well
described by a power-law model. Probably 4U\,0728$-$25 belongs to the class
of quasi-persistent sources that never reach a true quiescent state.
Also the long spin period of the pulsar makes it a good candidate for
accretion from the cold disc. The
unknown strength of the magnetic field of the NS in this system makes
estimates of the  propeller limiting luminosity impossible, which should
be much less than the observed luminosity in the case of a standard field.

\subsection{RX\,J0812.4$-$3114}

Bright type II outbursts were never observed from this source, however, a
sequence of normal (type I) outbursts occurred in December 1997
\citep{2000ApJ...530L..33C}. The {\it RXTE}/ASM
and {\it Swift}/BAT did not detect the source during the whole covered period. The {\it Chandra} observation revealed a thermal
spectrum with a very low temperature of $\sim0.1$ keV, which is the lowest
temperature among our sources. However, the very low counting statistics does not
allow us to firmly constrain the spectral parameters of this source. The observed luminosity $\sim2\times10^{33}$~\lum\ agrees with the estimates from the deep-crustal heating model.

\subsection{GS\,0834$-$430}

The {\it Chandra} observation utilized in this work was performed slightly
less than one year after the bright outburst of this source that occurred
in mid-2012. No other significant activity from the source is seen in its
light curve starting from 1996 (see Fig. \ref{fig:lcs}). The low number of
counts detected from this source does not allow us to discriminate
unambiguously between the power-law or blackbody spectral models. However, the
quite high blackbody temperature of about 2 keV and small radius of the emitting region hint to a non-thermal
origin of the spectrum. At the same time, the observed luminosity is very low
($\sim3\times10^{32}$ \lum). That means that either the crust is not heated
at all or the crust cooling time is very short ($<1$ year).

\subsection{GRO\,J1008$-$57}

The source was observed with {\it Chandra} four months after a giant type II
outburst and only two weeks after the type I outburst in mid-2013. Both the
high source luminosity ($\sim10^{35}$ \lum) and its spectral shape
(power law) point to a non-thermal origin of the emission. This conclusion is
supported by our recent monitoring of GRO\,J1008$-$57 with the {\it
Swift}/XRT that points to the transition of the source to the regime of
quasi-stable accretion from the cold disc \citep{tsyg2017}.

\subsection{2S\,1417$-$624}

{\it Chandra} observed 2S\,1417$-$624 about $3.3$ years after its bright
outburst that occurred in 2009 November. The source spectrum can be equally
well fitted with either a power-law or blackbody model with a quite high
temperature around 1.5 keV. A long time gap between the outburst and
observation would suggests the crust of the NS could have been in the
equilibrium with the core. However, then the high observed temperature (when
using the blackbody model) is unexpected (unless the cooling indeed
originate from very small hot spots of only 50 m in radius; Table 2),
pointing to a possibility of continuing accretion on to the NS at
a low level. Similar to MXB\,0656$-$072, this accretion can come from the cold
disc depleted after 2009 outburst.

\subsection{2S\,1553$-$542}

This source exhibited only three known outbursts in the history of X-ray observations (in 1975, 2007, and
2015). The {\it Chandra} data used in this paper were obtained approximately
5 years after the second outburst in 2007--2008. Taking into account that
2S\,1553$-$542 is the most distance source (20 kpc) in our sample
\citep{2016MNRAS.457..258T,2016MNRAS.462.3823L}, its counting statistics is very low and
prevents us from distinguishing between power-law and blackbody models.
However, the temperature obtained using the latter model is $\sim0.7$ keV,
and could point to the thermal origin of the emission that
can possibly be attributed to the NS surface emission. The observed luminosity
is much lower in comparison to the propeller limiting luminosity supporting this
conclusion.

\subsection{Swift\,J1626.6$-$5156}

The source was observed with the {\it Chandra} observatory twice: 3 and 7
years after the only bright outburst known for this source (occurring at the
beginning of 2006). Both observations show likely a non-thermal spectrum, and
no direct crust emission is observed. Despite that the luminosity has
dropped by a factor of $\sim4$ between the two observations, the spectral
parameters are very similar and consistent within the uncertainties. It is
worth mentioning that the expected limiting luminosity for the transition to
the propeller regime is roughly in the middle between the luminosities of
these two observations. Such behaviour agrees with the continuous accretion
from the cold disc.

Accretion during the brightest observation
is supported by the detection of the pulsations with a period of 15.3360(6) s
and a pulsed fraction of $54\pm7$ per cent.  In the second observation the
pulsations could not be detected and only an upper limit on the pulsed
fraction $\sim60$ per cent was obtained. This limit is consistent with the detected
value during the first observation and this non-detection is likely connected
with the much shorter exposure of this observation. The pulse period and
pulsed fraction values  obtained for the first {\it Chandra} observation
are consistent with values published by \cite{2011MNRAS.415.1523I} and are
typical for this source for periods of its flaring activity
\citep{2008A&A...485..797R}.

\subsection{GS\,1843+00}

The large distance to the source (10--15 kpc) and its low luminosity resulted
in very low counting statistics that does not allow us to draw firm
conclusions about the nature of the observed emission. The {\it Chandra}
observation was performed around 3 years after the last outburst detected in the {\it Swift}/BAT data. The observed luminosity is slightly lower
than the propeller limit because the source with such a long spin period likely accretes from the cold disc.

\subsection{XTE\,J1946+274}

The source showed a series of strong outbursts in 2011 April. Two years
later we measured using our {\it Chandra} data a source luminosity of about
$10^{34}$ \lum, which is significantly lower than the expected limiting
luminosity for the propeller regime onset (see Table \ref{tablum}). However,
we detected a quite hard spectrum (photon index of $\simeq1.1$)
and pulsations suggesting that we observe ongoing accretion on to the
NS surface. The spin period was measured to be 15.760(3) s and the pulsed
fraction $55\pm17$ per cent typical for the low-luminosity state (see Section~\ref{sec:time}).
Our spectral and temporal parameters agree with results of an independent
analysis of the same data performed by \cite{2015A&A...582A..53O}. The most probable explanation
of the hard spectrum of XTE\,J1946+274 below the propeller line is accretion from the
cold disc.

\subsection{KS\,1947+300}

Using {\it Chandra} we observed KS\,1947+300 about 6 years after the intense
flaring activity seen in 2001--2006 and nine months before another bright outburst in 2013 October
(Fig. \ref{fig:lcs}). The source was about
10 times brighter than the expected threshold luminosity for the propeller
regime onset and exhibited pulsations with a high pulsed fraction
($45\pm12$ per cent). In combination with the hard spectrum this indicates ongoing
accretion on to the NS surface probably from the cold disc.

\subsection{SAX\,J2103.5+4545}

The source was observed twice with {\it Chandra} during the same quiescent
period (in the middle of 2013 May and September). The results from the second
observation were published by \citet{2014MNRAS.445.1314R} and pointed to the
source being in a deep X-ray quiescence.
 We found similar source flux and spectral
parameters in both observations. As mentioned above when describing the
spectral fitting we fixed the absorption at the interstellar value, whereas
\citet{2014MNRAS.445.1314R} determined it from the spectral fitting  as
$N_{\rm H}=(0.3\pm0.1)\times10^{22}$ cm$^{-2}$. Fixing the absorption
parameter in both observations at this value results in blackbody model
parameters that are similar within the errors. We also detected pulsations in
the X-ray flux with comparable pulsed fractions of $76\pm25$ and $55\pm8$ per cent
for the first and second observations, respectively. It is interesting
that a similar high pulsed fraction (although with the different pulse shape)
was reported also for this source using an {\it XMM-Newton} observation when
the luminosity was three orders of magnitude larger
\citep{2004ApJ...616..463I}.

There are at least two ways to explain the observed results: (1) the crust did
not heat up much during the previous outburst and the NS crust was already in
equilibrium with the core during the first {\it Chandra} observation and
therefore the temperature did not decrease much further when second
observation was taken. If indeed cooling emission is observed then the
cooling in this system goes through hot spots \citep[such asymmetric cooling
might be possible for the NSs in BeXRPs, see discussion
in][]{2016MNRAS.463L..46W}; (2) in the other scenario we deal with a rather
stable low mass accretion rate on the NS surface. In this case, pulsations can
be due to accretion on to the magnetic poles, e.g. from the cold disc expected
to exist around the NS in this source.

\subsection{Cep\,X-4}

The {\it Chandra} observation was done almost 4 years after the 2009 March
outburst. However, the source exhibited low level flux variability even
between major flares explaining the clearly non-thermal spectrum and
significantly detected pulsations (pulsed fraction of $70\pm15$ per cent). Most
probably the observed power-law luminosity originates from the accretion from the
cold disc.

\subsection{SAX\,J2239.3+6116}

The source was discovered as a relatively faint transient source and never
exhibited strong type II outbursts. The quality of the available data does
not allow us to discriminate between the power-law or blackbody spectral models.
However, the observed luminosity around $10^{33}$ \lum\ is typical for BeXRPs
in a quiescence and the crust of the NS is likely in the equilibrium with the
core. Note that we do not show in Fig.\,\ref{fig:lcs} its light curves because
the source was not detected in the {\it Swift}/BAT and {\it RXTE}/ASM
data.

%%%%%%%%%%%%%%%%%%%%%%%%%%%%%%%%%%%%%%%%%%%%%%%%%%%%%%%%%%%%%%%%%%%%%%%%%%%%
%% DISCUSSION AND/OR CONCLUSIONS                                            %%
%%%%%%%%%%%%%%%%%%%%%%%%%%%%%%%%%%%%%%%%%%%%%%%%%%%%%%%%%%%%%%%%%%%%%%%%%%%%%%
\section{Discussion}
\label{discussion}

The observed luminosity in the quiescent state of BeXRPs potentially can be
explained with several emission mechanisms or a combination of them: (i)
magnetospheric accretion in the propeller regime, (ii) continuing
accretion on to the NS from the cold disc,
(iii) cooling of the NS surface that is heated due
to the accretion of matter when the source was in outburst, (iv) coronal activity
of the companion star. Below we discuss those possible explanations of the
detected emission and speculate about the role of the NS magnetic field.

%=================================================================
\begin{figure}
\includegraphics[width=\columnwidth, bb=70 145 500 710,clip]{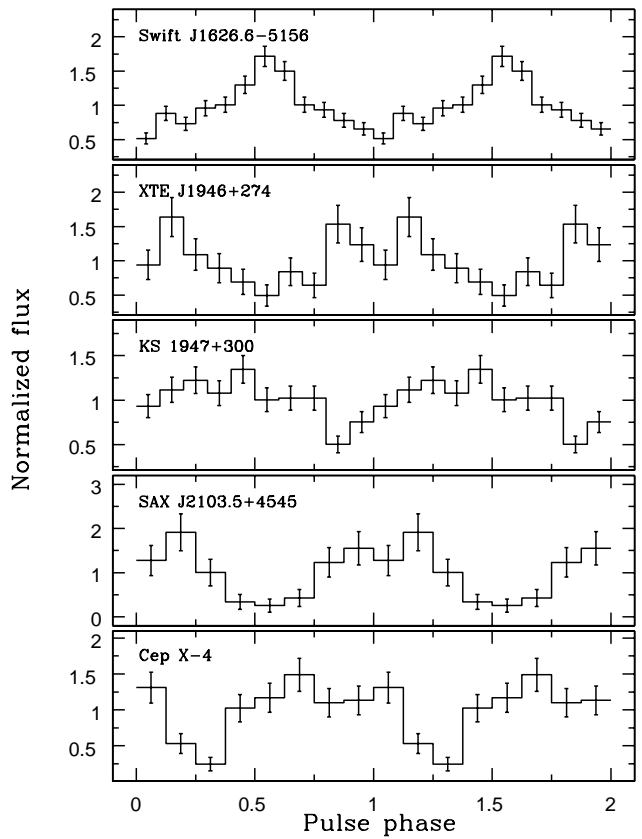}
\caption{Pulse profiles of five sources with detected pulsations
in the 0.5--10 keV energy range.}\label{fig:pprof}
\end{figure}
%=================================================================

\subsection{Centrifugal inhibition of accretion}
\label{prop}

The matter from the accretion disc can be accreted by the NS
only if the velocity of the magnetic field lines at the magnetospheric
radius is lower than the local Keplerian velocity. Otherwise the
pulsar enters the propeller regime and accretion will be
inhibited \citep{1975A&A....39..185I}. An abrupt drop of
the luminosity associated with the propeller regime onset was observed
in a number of accreting NSs with magnetic fields from $10^8$
to $10^{14}$ G \citep[see e.g.][and references
  therein]{2015SSRv..191..293R,2016MNRAS.457.1101T,2017ApJ...834..209L}.

The threshold value of the accretion
luminosity for the onset of the propeller can be roughly estimated by
equalising  the co-rotation radius $R_{\rm c}$ to the radius of the
magnetosphere $R_{\rm m}$ \citep[see e.g.][]{2002ApJ...580..389C}
\begin{multline}\label{eq1}
%\be\label{eq1}
L_{\rm prop}(R) = \frac{GM\dot{M}_{\rm prop}}{R} \\
\simeq 4 \times 10^{37} k^{7/2} B_{12}^2 P^{-7/3} M_{1.4}^{-2/3} R_6^5 \,\textrm{erg s$^{-1}$}
%\ee
\end{multline}
where $P$ is pulse period in seconds, $M_{\rm 1.4}$ is NS mass in units of
1.4M$_{\rm \odot}$, $R_6$ is NS radius in units of $10^6$ cm,
$B_{12}$ is magnetic field strength in units of $10^{12}$ G and $k$
is factor that accounts for the accretion flow geometry ($k\approx1$
in the case of spherically symmetric accretion and $k\approx0.5$ in the case
of disc accretion; \citealt{GL1978}).

The propeller effect does not permit the matter to penetrate into the
magnetosphere and to release its gravitational energy at the NS
surface. However, even under such circumstances one can expect to
detect an emission originating from the magnetosphere at the radius of $R_{\rm m}$
\citep{1996ApJ...457L..31C}. Therefore, the maximal luminosity in the
propeller regime could be achieved if $R_{\rm m}=R_{\rm c}$  \citep[e.g.][]{2002ApJ...580..389C}:
\begin{multline}\label{eq2}
%\be\label{eq1}
L_{\rm prop}(R_{\rm c}) = \frac{GM\dot{M}_{\rm prop}}{R_{\rm c}} \\
\simeq 2.4 \times 10^{35} k^{7/2} B_{12}^2 P^{-3} M_{1.4}^{-1} R_6^6 \,\textrm{erg s$^{-1}$}
%\ee
\end{multline}
A further increase of the mass accretion rate will shrink the magnetosphere
inside the co-rotation radius, allowing accretion on to the NS
surface to proceed. This will cause a sharp increase of the observed
luminosity.

It should be noted that the luminosity estimations done using equation
(\ref{eq2}) do not take into account the exact spectrum that would arise from
the magnetospheric accretion. It was shown by
\citet{1994ApJ...423L..47S} and \citet{2016A&A...593A..16T} that for NSs with a
strong magnetic field ($\sim10^{12}$ G) the accretion disc is disrupted at
large distances and that for a typical mass accretion rate of $10^{16}\,{\rm
g\,s^{-1}}$ one can expect a disc temperature of $\sim 10\,{\rm eV}\approx
10^5\,{\rm K}$. At such a temperature the bulk of the radiation should be
emitted in the UV band but not in the X-rays.

In contrast to $L_{\rm prop}(R_{\rm c})$ the limiting luminosity $L_{\rm prop}(R)$ calculated from
equation (\ref{eq1}) is not expected to be strongly affected by any
systematic biases and can be compared to the observations. The $L_{\rm prop}(R)$
for all sources in our sample that have known distances and magnetic field
strengths are listed in Table\,\ref{tablum} (assuming disc accretion so that
$k$=0.5). As can be seen, the majority of sources (with exception of
MXB\,0656$-$072, GRO\,J1008$-$57 and KS\,1947+300) emits well below the threshold
luminosity $L_{\rm prop}(R)$ for the propeller regime onset. Moreover, for
several sources we clearly detected the pulsations (see Table \ref{tablum})
demonstrating the compactness of emitting regions.

The fact that the observed emission level is below the critical one for the
propeller regime onset, but the pulsations are still detected, illustrates our poor
knowledge of what is going on in this regime. One of possible
explanations of the observed properties is a leakage of the matter
through the centrifugal barrier.  For instance it may happen via a so-called ``dead accretion
disc'' \citep{1977SvAL....3..138S} as a natural
reservoir of the matter around the NS under the condition of
the propeller state. If some low-level supply of the matter at the outer radius of the
accretion disc continues in the quiescent state one can expect
intermittent or even steady-state penetration of the matter through
the centrifugal barrier at the inner disc radius \citep[see
  e.g.][]{2010MNRAS.406.1208D,2012MNRAS.420..416D,2013A&A...550A..99Z}.
Another explanation proposed recently is a concept of stable accretion from
the cold recombined disc around the NS \citep{tsyg2017}, which we further explore in the next section.

\subsection{Stable accretion from the cold disc}
\label{sec:cold}

Until recently it was conventionally believed that all X-ray pulsars transit to the propeller regime once their luminosity is dropped to the value $L_{\rm prop}$ determined by  equation~(\ref{eq1}). However, \cite{tsyg2017} have shown that such transitions can be observed only in the sources where NSs rotate faster than some critical value for a given magnetic field strength. Otherwise a pulsar will start to accrete stably from a cold recombined accretion disc.

The main idea of the model is that the centrifugal barrier caused by the slowly rotating magnetosphere is strongly suppressed leading to a very low limiting luminosity $L_{\rm prop}$. For a certain combination of the pulsar's spin period and magnetic field strength, this limiting luminosity could be below the one (due to the low mass accretion rate) that would maintain the temperature throughout the accretion disc above the hydrogen recombination limit of $\sim 6500$ K. Therefore, the hydrogen in the disc would recombine and a cold accretion disc would be formed.

According to the thermal-viscous instability model this would happen in the case when the mass accretion rate drops below some critical value
\citep{1997ASPC..121..351L}:
\be\label{eq:Lasota}
  \dot{M}<\dot M_{\rm cold} \simeq 3.5\times 10^{15}\, r_{10}^{2.65}\,M_{1.4}^{-0.88}~~~{\rm g\,s^{-1}},
\ee
where $r_{10}=r/10^{10}\,{\rm cm}$ is the inner disc radius.

For a typical X-ray pulsar with a magnetic field around $2\times10^{12}$~G the corresponding critical luminosity is $L_{\rm cold}\sim3\times10^{34}$~\lum\ \citep{tsyg2017}. If during the outburst decay phase  this luminosity is reached before the propeller limiting
luminosity $L_{\rm prop}$, which is the case for all pulsars with the spin period longer than $\sim15$ s, the source will stop its fading and will transit to stable accretion from an entirely recombined cold disc.

It is very important to note here that X-ray pulsars that show this transition to accretion from the cold disc will never switch to the propeller regime because of the {\it decrease} of the inner radius of the disc as the source fades in contrast to accretion from a standard accretion disc where the inner radius is determined by the equation for the magnetospheric radius, i.e. it {\it increases} as the source fades. If no additional inflow is happening, which is probably true for the BeXRPs in between periastron passages, the luminosity will decrease with time as $L \propto t^{-0.7}$ \citep{tsyg2017}. Assuming $L_{\rm cold}\sim10^{35}$~\lum\ and absence of any mass inflow for a long time (few orbital cycles) then the pulsar can be found at any (low) luminosity.

Presence of the  pulsations is expected in the cold disc model because the inner disc radius  has always local (effective) temperature around $6500$ K, that allows the matter to interact with the magnetic field guiding it to the NS poles. It is worth noting that even the spectral properties
cannot give enough information to unambiguously determine whether the emission is of
thermal origin or is generated by the low-level accretion.
Indeed, as it was discussed by \cite{2016A&A...593A..16T} the influence of the
Comptonization of the X-ray spectrum, originated from the base of the accretion column,
by electrons in the optically thin atmosphere is negligible at the
accretion luminosity below $\sim 10^{34}-10^{35}\,{\rm erg\,s^{-1}}$
because the Thomson optical thickness across the accretion channel is
well below unity \citep{2015MNRAS.447.1847M}. However, the shape of the initial spectrum depends
on the braking distance of the accretion flow, which might be affected by MHD waves,
and the opacity in the NS atmosphere. This problem is still under investigation.

\subsection{Deep-crustal heating}
\label{sec:heating}

Another possible source of the emission from quiescent BeXRPs is the
surface of the NS that is cooling down after being heated up
in the outbursts by the  accreted matter. This requires a
substantial amount of matter to be accumulated on the NS surface in
the course of an outburst. This settling matter compresses the
original NS crust and enriches it with low-Z elements
\citep{1990A&A...229..117H}. It gives rise to non-equilibrium
reactions (electron captures, neutron emissions and pycnonuclear
reactions; see e.g. \citealt{1990A&A...227..431H,
  2003A&A...404L..33H, 2007ApJ...662.1188G, 2008A&A...480..459H,
  2012PhRvC..85e5804S}), although the main source of heat generation
are the pycnonuclear reactions proceeding deep in the crust at high
densities ($>10^{12}$ g cm$^{-3}$). Most of this heat is conducted
inwards to the core, which looses its energy via neutrino emission,
but a small part is radiated from the NS surface
as thermal photons. This  ''deep-crustal heating'' \citep{1998ApJ...504L..95B}
will slowly heat up the
crust.  Through this process the long-term averaged mass accretion rate onto
the NS is connected to the surface thermal emission (which is
a direct probe for the NS core temperature). Particularly,
an averaged accretion rate of $\langle \dot{M} \rangle= 10^{-11}$ M$_{\odot}$ yr$^{-1}$
will provide an averaged thermal NS
luminosity at a level of about $6\times10^{32}$
\ergs.

We note that this assumes that only standard, slow, neutrino cooling
processes (e.g. modified Urca, bremsstrahlung) occurring in the core. If
fast neutrino cooling (e.g. direct Urca or the neutrino processes that arise
if exotic material like mesons, hyperons or unbound quarks are present in the core)
occur in the core, then the thermal luminosities from the surface might be
significantly reduced \citep[see][]{1998ApJ...504L..95B,2004ARA&A..42..169Y,
2009ApJ...691.1035H, 2010ApJ...714..894H, 2013MNRAS.432.2366W}. However, we
note that those enhanced core neutrino processes mostly become active when
the NSs are relatively massive \citep[][]{2001ApJ...548L.175C}.
Since the NSs in BeXRPs might not have accreted a large amount of matter yet
(because of the relatively young age of those systems), it is unclear if
indeed enhanced core neutrino processes can be activated in the NSs in
BeXRPs.

The averaged mass accretion rate is thus the crucial parameter to
estimate the luminosity of the cooling NS predicted by the
deep-crustal heating model. Most of the classical BeXRPs spend the
majority of their life in quiescence, showing bright outbursts only
from time to time.  Unfortunately, there is in general no strict
recurrence time (except for those that show recurrent type-I outburst
every periastron passage) between periods of the activity of Be
companions and, hence, the only way to estimate an averaged mass
accretion rate is direct observations. In this paper, we used for these
estimations data from the {\it RXTE}/ASM and {\it SWIFT}/BAT monitors
covering together an $\sim19$ years time-span (1996--2015). To
calculate an averaged flux we followed the procedure described by
\cite{2016MNRAS.463...78E}. Namely, in the 2--10 keV energy range the
averaged count rate obtained from the {\it RXTE}/ASM data was
converted to the flux using the Crab spectrum with standard parameters
(photon index of 2.1 and normalization of 10 ph keV$^{-1}$ cm$^{-2}$ s$^{-1}$ at 1 keV), and above 15 keV the
flux was derived the same way but using the {\it Swift}/BAT data. Taking into account difference between spectra of the
Crab nebula and typical XRP \citep[we adopted the cut-off power-law model with photon index $\Gamma=1.0$ and cut-off energy $E_{cut}=15$ keV; ][]{2005AstL...31..729F}, the resultant flux must be reduced by $\sim30$\%.
We translated it into a luminosity and, consequently, to
the mass accretion rate using the simple equation $\langle \dot{M}
\rangle=L \,R /G \,M$ (see Table
\ref{tablum}). The expected quiescent luminosities from the
deep-crustal heating were calculated as $L_{\rm q}= (\langle
\dot{M}\rangle  / 10^{-11}\mbox{M}_{\odot} \mbox{yr}^{-1})   \times 6\times10^{32}$ \lum\ and are presented
in the same table. It is worth noting that due to a very limited
history of observations our estimations should be considered as very
rough ones with an accuracy not better than an order of magnitude.

Another potential problem is that due to a relatively low sensitivity of
the above-mentioned all-sky monitors the calculated averaged mass accretion
rates are mainly defined by the source fluxes during their outburst activity
while the fluxes below the detection threshold are ignored. In the case of
the pulsars able to transit to the propeller regime one may not expect significant heating during
quiescence since accretion would be inhibited. However, as it follows from the cold disc model and
our analysis, the low-level accretion can still proceed for the sources with relatively long spin periods.
Thus, the averaged mass accretion rate
estimated using only outbursts activity periods should be treated as a
lower limit. However, even if the observed quiescent luminosity in all
our sources is due to accretion and always has the same level in
quiescence, it will not shift our estimations significantly.  For
instance, a luminosity of $1\times10^{34}$ \ergs\ corresponds to a mass
accretion rate of about $9\times10^{-13}$ M$_{\odot}$ yr$^{-1}$, which
is at least two orders of magnitude lower than the values
obtained from the all-sky monitors data (see Table \ref{tablum}).

Some pulsars from our sample have observed luminosity $L$ well above
predicted $L_{\rm q}$ (e.g. MXB\,0656$-$072, 4U\,0728$-$25,
GRO\,J1008$-$57, Swift\,J1626.6$-$5156, XTE\,J1946+274,
KS\,1947+300). Most of these have also hard X-ray spectra and strong
X-ray pulsations. Taken together, this suggests that accretion is
ongoing in these systems (most probably from the cold disc). Other
pulsars, on the other hand, have an observed luminosity that is
comparable or lower than the quiescent thermal luminosity predicted to
arise from deep crustal heating based on the observed outburst
properties (e.g. 4U\,0115+63, V\,0332+53, RX\,J0812.4$-$3114,
GS\,0834$-$430, 2S\,1417$-$624, 2S\,1553$-$542, GS\,1843+00,
SAX\,J2103.5+4545). The first three of these also exhibit soft thermal
spectra. This may suggest that we observe thermal emission from the
NS in these objects.  For some sources the observed
luminosity is significantly lower than predicted from the deep crustal
heating model. There could be several reasons for this discrepancy (see also the discussion in \cite{2016MNRAS.463...78E} with respect to V\,0332+53). We could have significantly overestimated the long-term averaged mass accretion rates. This could happen if the outburst histories of our sources over the past 19 years are not representative for their activity over the last few millennia (the time-scale that governs the thermal evolution of the core; \citealt{2001ApJ...548L.175C,2013MNRAS.432.2366W}). If in the past the sources were less active the averaged accretion rates would be significantly lower and the estimated quiescent luminosities would become more comparable with those observed.

Alternatively, the NS cores could cool significantly faster than assumed in the standard model, potentially indicating the possible presence of enhanced cooling processes in the core. This would require relatively massive NSs \citep{2004ARA&A..42..169Y}. Although some NSs might have been born massive \citep[see the discussion in ][]{2016MNRAS.463...78E}, it is unclear if this can be true for all of our cooling candidates.

Another possibility is that the heating in the crusts of those sources
is less efficient than assumed.  As it was discussed by
\cite{2013MNRAS.432.2366W} a substantial amount of matter has to be
accreted before the original NS crust would be replaced with the new
one, that would be enriched with low-Z element. In order to accrete
$10^{-4}$ M$_{\odot}$, the estimated averaged mass accretion rates
(see Table \ref{tablum}) have to be constant during
$\sim10^{5}$--$10^{7}$ years, that is comparable with the lifetime of
HMXBs, and therefore some NSs in our sample could have a crust that is
not yet fully replaced. It is currently unclear if/how the nuclear
heating of such a ``hybrid'' crust is different from that of a fully
accreted crust (which is assumed in standard heating models, e.g.
\citealt{2008A&A...480..459H}). If some of our targets have
such a hybrid crust then the deep crustal reactions might be
affected and maybe in some sources even inhibited.  Consequently,
those NSs might not have been heated significantly even if they would
accrete currently at a high averaged $\dot M$. Since we do not know
the exact composition of the NS crust in any system, this adds another
uncertainty in our interpretation of the data.

Finally,  when fitting the spectra of the thermal sources with a blackbody model, the radii of the emitting regions are very small, much smaller than the NS radius (this is also true if the spectra are fitted with a NS atmosphere model; see \citealt{2016MNRAS.463...78E}). This could indicate that the cooling occurs via hotspots at the magnetic poles \citep[see the discussion in ][]{2016MNRAS.463L..46W}. This would cause pulsations in the light curves but the quality of our data does not allow to put constraints on that possibility for those sources. However, as determined by \cite{2016MNRAS.463...78E} for V\,0332+53, the rest of the NS surface   also emits some radiation. Although the temperature of the rest of the surface should be significantly lower than what is observed from the hotspots, the emitting region is much larger and the total luminosity emitted from the rest of the NS surface could be significantly larger than the power emitted from the hotspots. Therefore, the sources could still be consistent with the standard heating and cooling models (see e.g. figure 3 of \citealt{2016MNRAS.463...78E}). Unfortunately, our current data sets do not allow us to put any constraints on the possible emission from the rest of the surface. However, this potential contribution has to be taken into account when modelling the cooling emission from high-magnetic field NSs.

Besides that we might directly probe the NS cores for our cooling candidates, it is possible that during some outbursts (i.e. the giant outbursts) enough heat was generated in the crust so that the crust departed from thermal equilibrium with the core. In such cases, the thermal luminosity would represent the thermal state of the crust and not that of the core and potentially we could observe (using multiple observations of the same source) the crust cooling until equilibrium is restored \citep[see ][ for this possibility]{2016MNRAS.463L..46W}. The source GS\,0834$-$430 was observed within 1 year after its type-II outburst in 2012 at a rather low luminosity of $3\times10^{32}$ \ergs. This low luminosity would indicate that the crust was already in thermal equilibrium with the core during the time of our observation, unless the core is very cold. If truly in equilibrium, this would constrain the crust cooling time-scale to be less than 1 year. This is consistent with what is reported by \cite{2017arXiv170400284R} who found a crust cooling time-scale of less than 250 days for 4U\,0115+63. All the other sources in our sample were observed much longer after their last giant outbursts. Despite that it is not clear that the crust cooling time-scales in all NSs should be of the same duration, it is plausible that for all our cooling candidates, the crust was in equilibrium with the core during the time of our observations.

\subsection{Contamination from the optical counterpart}

The majority of previous studies dedicated to the thermal emission of NSs
in the quiescent state were based on observations of LMXBs. However, in
contrast to LMXBs, BeXRPs harbour massive and hot optical companions of O and
B classes, which are known to emit some non-negligible fraction of their
bolometric luminosity in X-rays
\citep{1979ApJ...229..304C,1979ApJ...234L..55S}.

Systematic studies of X-ray properties of O and B stars have been done
using different observatories (see e.g. \citealt{1997A&A...322..167B}
for the {\it ROSAT} data, \citealt{2009A&A...506.1055N} for the {\it XMM-Newton}
data, \citealt{2011ApJS..194....7N} for the {\it Chandra} data, and references
therein). Particularly, it was shown that spectra of O stars can be
well fitted by a blackbody model with a temperature between $\sim$0.2
and $\sim$0.6 keV with additional absorption above the interstellar one.
The B stars have harder spectra with temperatures around 1 keV and
usually do not require additional local absorption.

The X-ray luminosity of O stars clearly correlates with the bolometric one,
whereas for B stars such a correlation is not obvious and has a large
scatter. Besides the intrinsic scatter, some additional one can be
introduced by the magnetic field of the star and its binary nature
(particularly, due to an interaction of the emission from the compact object
with the stellar wind).

Using fig.\,3 from \citet{2011ApJS..194....7N}, we can roughly
estimate the expected X-ray luminosities for the optical companions of the pulsars
in our sample. As can be seen from Table\,\ref{tabsam}, the spectral types of
companions cover a relatively narrow range from O8 to B2. According to the
abovementioned paper, such stars emit around $10^{-7}$ of their
bolometric luminosity in the X-rays, which makes several times $10^{31}$ \ergs.
This value is about two orders of magnitude lower than the luminosity
measured in our observations for the majority of sources. The lowest
luminosities are demonstrated by 4U\,0115+63 and V\,0332+53 and, at the same
time, those systems are amongst the one with the hottest optical companions.
Therefore, these sources are
the only objects from the sample where some contribution from the optical
star may be present but it is not expected to be significant.

\section{Conclusions}

In this work we presented results of the first systematic survey of the
sample of 16 X-ray pulsars with Be optical companions during quiescent
state. The observations were taken with the {\it Chandra}, {\it
  XMM-Newton} and {\it Swift} observatories with different delays
after an outburst activity of the sources. This gave us a possibility to
study thermal evolution of highly magnetized NSs in
details.  Surprisingly, a substantial fraction of the sources from
our sample have hard energy spectra even at low luminosities implying continuing accretion.
Five of them show strong
pulsations with pulse fraction of 50--70 per cent. Such behaviour can be
explained in terms of the recently proposed model of stable accretion
from the cold disc \citep{tsyg2017}. For the rapidly rotating pulsars that are
able to transit to the propeller regime much softer spectra were
observed. Measured blackbody temperature and very low luminosity
indicate cooling process of the NS heated up in the
outbursts by the accreted matter.  The obtained results were discussed
within the framework of the ``deep-crustal heating'' model \citep{1998ApJ...504L..95B}.

To make more strict conclusions on the cooling processes in the magnetized
NSs, deeper multiple observations of the pulsars in their quiescent state are
needed. Additionally, to be able to robustly discriminate thermal
spectra from the non-thermal ones, very deep observations should be performed with
instruments possessing a high sensitivity in a broad energy range, e.g. the
{\it NuSTAR} observatory. Furthermore, such observations should be performed
only for the sources that are able to transit to the propeller regime, i.e.
rotating rapidly enough to centrifugally halt the accretion before the disc
transits to the fully recombined state \citep{tsyg2017}.

%%%%%%%%%%%%%%%%%%%%%%%%%%%%%%%%%%%%%%%%%%%%%%%%%%%%%%%%%%%%%%%%%%%%%%%%%%%%%%
%% Acknowledgments                                                         %%
%%%%%%%%%%%%%%%%%%%%%%%%%%%%%%%%%%%%%%%%%%%%%%%%%%%%%%%%%%%%%%%%%%%%%%%%%%%%%%
\section*{Acknowledgements}

We thank Alexander Mushtukov and Valery Suleimanov for helpful discussions.
This work made use of the {\it Chandra} and {\it XMM-Newton} public data archives as well as
data of {\it Swift} supplied by the UK Swift Science Data Centre at
the University of Leicester.
AAL and SST acknowledge support by the Russian Science Foundation grant 14-12-01287 in part of the individual sources analysis and theoretical discussions (Sections 3 and 4). AAL also thanks the Dynasty Foundation for its support in analysing of the X-ray data (Section 2).
RW is supported by a NWO Top Grant, Module 1.
ND acknowledges support via an NWO/Vidi grant and an EU Marie Curie Intra-European fellowship under contract no. FP-PEOPLE-2013-IEF-627148.
JP was partially supported by the Academy of Finland grant 268740 and  the National Science Foundation grant PHY-1125915.

%%%%%%%%%%%%%%%%%%%%%%%%%%%%%%%%%%%%%%%%%%%%%%%%%%%%%%%%%%%%%%%%%%%%%%%%%%%%%%
%% Bibliography                                                             %%
%%%%%%%%%%%%%%%%%%%%%%%%%%%%%%%%%%%%%%%%%%%%%%%%%%%%%%%%%%%%%%%%%%%%%%%%%%%%%%

\bibliographystyle{mnras}
\bibliography{allbib}

%=================================================================
\begin{landscape}
\vspace{-20mm}\begin{table}
\noindent
\tiny
\centering
\caption[Spectral parameters]{Spectral parameters} \label{tab:spec_all}
\begin{tabular}{l|c|c|c|c|c|c|c|c|c|c|c}

\hline\hline
&&&&&&&&&&&\\
Source  & ObsID &  Date & Exposure  & $N_{\rm H}$ & $kT_{\rm bb}$  & $R_{\rm bb}$ & $\Gamma$ & Observed flux (0.5--10 keV)  & Unabsorbed flux (0.5--10 keV)  & $L_{0.5-10\ \rm{keV}}$ & C-value (dof) \\[1mm]
        &       &  (MJD)   &  (ks)     & ($10^{22}$\ cm$^{-2}$)                 & (keV)      &  (km)     &             & (erg\,s$^{-1}$ cm$^{-2}$)                 &     (erg\,s$^{-1}$ cm$^{-2}$)              &  (erg\,s$^{-1}$)  &              \\
\hline
&&&&&&&&&&&\\
4U\,0115+63 & 0505280101$^{a}$ & 54302.07 & 13.3 & 0.86 & 0.31$^{+0.03}_{-0.02}$  & 0.76$^{+0.19}_{-0.15}$ & -- & 2.93$^{+0.31}_{-0.40}\times10^{-14}$ & 1.00$^{+0.12}_{-0.11}\times10^{-13}$ &  5.86$^{+0.72}_{-0.64}\times10^{32}$ & 161.7(188) \\[1mm]
 &  &  &  & 0.86 & -- & -- & 3.5$^{+0.3}_{-0.3}$ & 3.85$^{+0.53}_{-0.48}\times10^{-14}$ & 1.95$^{+0.34}_{-0.25}\times10^{-13}$ &  1.14$^{+0.20}_{-0.15}\times10^{33}$ & 171.1(188) \\[2mm]

V\,0332+53 & 1919 & 51913.17 & 5.1 & 0.70  & 0.38$^{+0.09}_{-0.07}$  & 0.32$^{+0.21}_{-0.12}$ & -- & 1.93$^{+0.51}_{-0.51}\times10^{-14}$ & 4.37$^{+1.13}_{-1.05}\times10^{-14}$ &  2.56$^{+0.66}_{-0.62}\times10^{32}$ & 14.6(12) \\[1mm]
 &  &  &  & 0.70 & -- & -- & 3.2$^{+0.7}_{-0.6}$ & 2.55$^{+0.69}_{-0.72}\times10^{-14}$ & 9.12$^{+4.68}_{-2.66}\times10^{-14}$ &  5.35$^{+2.75}_{-1.56}\times10^{32}$ & 12.7(12) \\[2mm]

 & 0506190101$^{a}$ & 54506.95 & 14.2 & 0.70 & 0.43$^{+0.05}_{-0.04}$  & 0.29$^{+0.08}_{-0.06}$ & -- & 2.88$^{+0.41}_{-0.34}\times10^{-14}$ & 5.75$^{+0.70}_{-0.74}\times10^{-14}$ &  3.37$^{+0.41}_{-0.44}\times10^{32}$ & 124.0(155) \\[1mm]
 &  &  &  & 0.70 & -- & -- & 2.7$^{+0.3}_{-0.3}$ & 3.91$^{+0.62}_{-0.52}\times10^{-14}$ & 1.00$^{+0.15}_{-0.13}\times10^{-13}$ &  5.86$^{+0.87}_{-0.76}\times10^{32}$ & 124.0(155) \\[2mm]

MXB\,0656$-$072 & 14635 & 56278.55 & 4.6 & 0.60 & 0.97$^{+0.04}_{-0.04}$  & 0.18$^{+0.01}_{-0.01}$ & -- & 1.78$^{+0.11}_{-0.14}\times10^{-12}$ & 2.09$^{+0.15}_{-0.14}\times10^{-12}$ &  3.80$^{+0.27}_{-0.25}\times10^{33}$ & 213.0(284) \\[1mm]
 &  &  &  & 0.60 & -- & -- & 1.2$^{+0.1}_{-0.1}$ & 2.61$^{+0.18}_{-0.15}\times10^{-12}$ & 3.09$^{+0.22}_{-0.21}\times10^{-12}$ &  5.62$^{+0.40}_{-0.38}\times10^{33}$ & 221.5(284) \\[2mm]

4U\,0728$-$25 & 0003800500N$^{b}$ & 54665.59 & 17.3 & 0.54 & 1.00$^{+0.05}_{-0.04}$  & 0.25$^{+0.02}_{-0.02}$ & -- & 1.56$^{+0.10}_{-0.08}\times10^{-12}$ & 1.78$^{+0.13}_{-0.08}\times10^{-12}$ &  7.91$^{+0.57}_{-0.36}\times10^{33}$ & 276.4(282) \\[1mm]
 &  &  &  & 0.54 & -- & -- & 1.3$^{+0.1}_{-0.1}$ & 2.09$^{+0.13}_{-0.16}\times10^{-12}$ & 2.51$^{+0.12}_{-0.17}\times10^{-12}$ &  1.12$^{+0.05}_{-0.07}\times10^{34}$ & 239.7(282) \\[2mm]

 & 00038005004$^{b}$ & 54802.36 & 2.2 & 0.54 & 1.33$^{+0.50}_{-0.29}$  & 0.10$^{+0.05}_{-0.04}$ & -- & 7.90$^{+3.01}_{-2.68}\times10^{-13}$ & 8.71$^{+3.59}_{-2.54}\times10^{-13}$ &  3.88$^{+1.60}_{-1.13}\times10^{33}$ & 13.3(18) \\[1mm]
 &  &  &  & 0.54 & -- & -- & 0.8$^{+0.5}_{-0.5}$ & 1.07$^{+0.54}_{-0.31}\times10^{-12}$ & 1.20$^{+0.42}_{-0.33}\times10^{-12}$ &  5.35$^{+1.87}_{-1.47}\times10^{33}$ & 11.5(18) \\[2mm]

RX\,J0812.4$-$3114 & 14637 & 56486.52 & 4.6 & 0.48 & 0.13$^{+0.03}_{-0.02}$  & 10.0$^c$ & -- & 1.92$^{+0.55}_{-0.56}\times10^{-14}$ & 1.58$^{+0.81}_{-0.61}\times10^{-13}$ &  1.60$^{+0.82}_{-0.62}\times10^{33}$ & 8.3(18) \\[1mm]
 &  &  &  & 0.48 & -- & -- & 5.6$^{+1.1}_{-1.1}$ & 2.37$^{+0.88}_{-0.37}\times10^{-14}$ & 2.51$^{+1.56}_{-1.03}\times10^{-13}$ &  2.54$^{+1.58}_{-1.05}\times10^{33}$ & 8.5(18) \\[2mm]

GS\,0834$-$43 & 14638 & 56478.08 & 4.6 & 1.00 & 1.92$^{+4.04}_{-0.72}$  & 0.01$^{+0.01}_{-0.01}$ & -- & 8.23$^{+6.48}_{-3.63}\times10^{-14}$ & 8.91$^{+9.71}_{-4.01}\times10^{-14}$ &  2.67$^{+2.90}_{-1.20}\times10^{32}$ & 9.4(16) \\[1mm]
 & &  &  & 1.00 & -- & -- & 0.2$^{+0.8}_{-0.8}$ & 1.15$^{+0.74}_{-0.52}\times10^{-13}$ & 1.23$^{+0.91}_{-0.52}\times10^{-13}$ &  3.68$^{+2.71}_{-1.56}\times10^{32}$ & 9.3(16) \\[2mm]

GRO\,J1008$-$57 & 14639 & 56440.72 & 4.6 & 1.40 & 1.22$^{+0.02}_{-0.02}$  & 0.49$^{+0.02}_{-0.02}$ & -- & 1.36$^{+0.04}_{-0.04}\times10^{-11}$ & 1.66$^{+0.04}_{-0.04}\times10^{-11}$ &  6.68$^{+0.16}_{-0.15}\times10^{34}$ & 658.2(484) \\[1mm]
 &  &  &  & 1.40 & -- & -- & 1.05$^{+0.04}_{-0.04}$ & 1.84$^{+0.05}_{-0.05}\times10^{-11}$ & 2.34$^{+0.05}_{-0.05}\times10^{-11}$ &  9.43$^{+0.22}_{-0.21}\times10^{34}$ & 450.1(484) \\[2mm]

2S\,1417$-$624 & 14640 & 56432.32 & 4.6 & 1.30 & 1.57$^{+0.34}_{-0.24}$  & 0.05$^{+0.01}_{-0.01}$ & -- & 4.07$^{+1.20}_{-0.68}\times10^{-13}$ & 4.68$^{+1.08}_{-0.88}\times10^{-13}$ &  2.01$^{+0.46}_{-0.38}\times10^{33}$ & 80.9(68) \\[1mm]
& &  &  & 1.30 & -- & -- & 0.6$^{+0.3}_{-0.3}$ & 5.43$^{+0.99}_{-0.98}\times10^{-13}$ & 6.17$^{+1.25}_{-0.92}\times10^{-13}$ &  2.66$^{+0.54}_{-0.40}\times10^{33}$ & 76.84(68) \\[2mm]

2S\,1553$-$542 & 14641 & 56389.32 & 4.6 & 1.70 & 0.66$^{+0.65}_{-0.25}$  & 0.20$^{+0.56}_{-0.14}$ & -- & 1.18$^{+0.49}_{-0.44}\times10^{-14}$ & 2.19$^{+1.36}_{-0.93}\times10^{-14}$ &  1.05$^{+0.65}_{-0.44}\times10^{33}$ & 5.8(3) \\[1mm]
 &  &  &  & 1.70 & -- & -- & 2.4$^{+1.8}_{-1.4}$ & 1.71$^{+1.16}_{-0.99}\times10^{-14}$ & 4.68$^{+16.70}_{-2.28}\times10^{-14}$ &  2.24$^{+7.99}_{-1.09}\times10^{33}$ & 4.8(3) \\[2mm]

Swift\,J1626.6$-$5156 & 10049 & 54897.31 & 18.2 & 1.00 & 1.20$^{+0.04}_{-0.04}$  & 0.22$^{+0.01}_{-0.01}$ & -- & 8.94$^{+0.37}_{-0.42}\times10^{-13}$ & 1.05$^{+0.05}_{-0.05}\times10^{-12}$ &  1.25$^{+0.06}_{-0.06}\times10^{34}$ & 420.0(359) \\[1mm]
 &  &  &  & 1.00 & -- & -- & 1.05$^{+0.07}_{-0.07}$ & 1.19$^{+0.06}_{-0.07}\times10^{-12}$ & 1.45$^{+0.07}_{-0.07}\times10^{-12}$ &  1.73$^{+0.08}_{-0.08}\times10^{34}$ & 315.9(359) \\[2mm]

 & 14642 & 56307.97 & 4.6 & 1.00 & 1.22$^{+0.23}_{-0.17}$  & 0.10$^{+0.03}_{-0.02}$ & -- & 2.05$^{+0.54}_{-0.40}\times10^{-13}$ & 2.40$^{+0.55}_{-0.40}\times10^{-13}$ &  2.87$^{+0.66}_{-0.48}\times10^{33}$ & 53.4(46) \\[1mm]
 &  &  &  & 1.00 & -- & -- & 1.0$^{+0.3}_{-0.3}$ & 2.91$^{+0.85}_{-0.62}\times10^{-13}$ & 3.47$^{+0.70}_{-0.58}\times10^{-13}$ &  4.15$^{+0.84}_{-0.70}\times10^{33}$ & 50.6(46) \\[2mm]

GS\,1843+00 & 14644 & 56452.32 & 4.6 & 1.00 & 2.20$^{+6.80}_{-0.92}$  & 0.03$^{+0.03}_{-0.03}$ & -- & 7.71$^{+8.50}_{-4.15}\times10^{-14}$ & 8.32$^{+7.17}_{-3.53}\times10^{-14}$ &  1.55$^{+1.34}_{-0.66}\times10^{33}$ & 8.1(7) \\[1mm]
 &  &  &  & 1.00 & -- & -- & 0.2$^{+1.1}_{-1.1}$ & 9.91$^{+6.55}_{-5.79}\times10^{-14}$ & 1.07$^{+0.71}_{-0.41}\times10^{-13}$ &  2.00$^{+1.32}_{-0.77}\times10^{33}$ & 8.6(7) \\[2mm]
\hline
\end{tabular}
\end{table}
\end{landscape}

\pagebreak

\begin{landscape}
\begin{table}
\noindent
\tiny
\centering
%\ContinuedFloat

\contcaption{Spectral parameters} \label{tab:spec_all2}
\begin{tabular}{l|c|c|c|c|c|c|c|c|c|c|c}

\hline\hline
&&&&&&&&&&&\\
Source  & ObsID &  Date & Exposure  & $N_{\rm H}$ & $kT_{\rm bb}$  & $R_{\rm bb}$ & $\Gamma$ & Observed flux (0.5--10 keV)  & Unabsorbed flux (0.5--10 keV)  & $L_{0.5-10\ \rm{keV}}$ & C-value (dof) \\[1mm]
        &       &  (MJD)   &  (ks)     & ($10^{22}$\ cm$^{-2}$)                 & (keV)      &  (km)     &             & (erg\,s$^{-1}$ cm$^{-2}$)                 &     (erg\,s$^{-1}$ cm$^{-2}$)              &  (erg\,s$^{-1}$)  &              \\
        %[1mm]
\hline
&&&&&&&&&&&\\
XTE\,J1946+274 & 14646 & 56363.12 & 4.6 & 0.90 & 1.11$^{+0.09}_{-0.08}$  & 0.21$^{+0.03}_{-0.02}$ & -- & 7.02$^{+0.88}_{-0.61}\times10^{-13}$ & 8.32$^{+0.80}_{-0.73}\times10^{-13}$ &  8.06$^{+0.78}_{-0.71}\times10^{33}$ & 122.8(148) \\[1mm]
 & & &  & 0.90 & -- & -- & 1.1$^{+0.2}_{-0.2}$ & 9.92$^{+1.05}_{-0.85}\times10^{-13}$ & 1.20$^{+0.12}_{-0.11}\times10^{-12}$ &  1.16$^{+0.11}_{-0.10}\times10^{34}$ & 108.0(148) \\[2mm]

KS\,1947+300 & 14647 & 56297.13 & 4.6 & 0.90 & 1.32$^{+0.07}_{-0.06}$  & 0.30$^{+0.02}_{-0.02}$ & -- & 2.43$^{+0.18}_{-0.13}\times10^{-12}$ & 2.75$^{+0.20}_{-0.12}\times10^{-12}$ &  3.29$^{+0.24}_{-0.15}\times10^{34}$ & 256.7(302) \\[1mm]
 &  &  &  & 0.90 & -- & -- & 0.8$^{+0.1}_{-0.1}$ & 3.42$^{+0.19}_{-0.24}\times10^{-12}$ & 3.89$^{+0.28}_{-0.18}\times10^{-12}$ &  4.65$^{+0.33}_{-0.21}\times10^{34}$ & 241.6(302) \\[2mm]

SAX\,J2103.5+4545 & 14648 & 56429.01 & 4.6 & 0.66 & 0.88$^{+0.09}_{-0.08}$  & 0.14$^{+0.02}_{-0.02}$ & -- & 2.41$^{+0.26}_{-0.30}\times10^{-13}$ & 2.95$^{+0.44}_{-0.32}\times10^{-13}$ &  1.49$^{+0.22}_{-0.16}\times10^{33}$ & 79.9(72) \\[1mm]
 &  &  &  & 0.66 & -- & -- & 1.3$^{+0.2}_{-0.2}$ & 3.89$^{+0.96}_{-0.49}\times10^{-13}$ & 4.79$^{+0.71}_{-0.62}\times10^{-13}$ &  2.42$^{+0.36}_{-0.31}\times10^{33}$ & 75.6(72) \\[2mm]

Cep\,X-4 & 14649 & 56344.08 & 4.4 & 0.80 & 0.88$^{+0.08}_{-0.07}$  & 0.13$^{+0.02}_{-0.02}$ & -- & 5.60$^{+0.72}_{-0.41}\times10^{-13}$ & 7.08$^{+0.86}_{-0.62}\times10^{-13}$ &  1.22$^{+0.15}_{-0.11}\times10^{33}$ & 108.6(153) \\[1mm]
 &  &  &  & 0.80 & -- & -- & 1.4$^{+0.2}_{-0.2}$ & 8.67$^{+0.95}_{-1.33}\times10^{-13}$ & 1.12$^{+0.11}_{-0.12}\times10^{-12}$ &  1.94$^{+0.19}_{-0.21}\times10^{33}$ & 101.0(153) \\[2mm]

SAX\,J2239.3+6116 & 14650 & 56361.59 & 4.6 & 0.85 & 1.62$^{+0.43}_{-0.30}$  & 0.04$^{+0.01}_{-0.01}$ & -- & 4.01$^{+0.91}_{-0.89}\times10^{-13}$ & 4.37$^{+1.26}_{-0.98}\times10^{-13}$ &  1.01$^{+0.29}_{-0.23}\times10^{33}$ & 56.5(83) \\[1mm]
 &  &  &  & 0.85 & -- & -- & 0.4$^{+0.3}_{-0.3}$ & 5.77$^{+1.47}_{-1.28}\times10^{-13}$ & 6.31$^{+1.45}_{-1.18}\times10^{-13}$ &  1.46$^{+0.34}_{-0.27}\times10^{33}$ & 56.4(83) \\[2mm]

\hline
\end{tabular}
\begin{flushleft}{
$^{a}$Observations performed by {\it XMM-Newton}  \\
$^{b}$Observations performed by {\it Swift}/XRT; ObdID. 0003800500N means the sum of 00038005001-00038005003 observations. \\
$^{c}$1$\sigma$ upper limit on the blackbody radius in RX\,J0812.4$-$3114.
  }\end{flushleft}
\end{table}
\end{landscape}
%=================================================================

%=================================================================================
%\begin{landscape}
\begin{table}
\noindent
\centering
\caption{Determined parameters of the studied BeXRPs.}\label{tablum}
\centering
\vspace{1mm}
%\small{
%\begin{minipage}{\textwidth}
  \begin{tabular}{|l|c|c|c|c|c|c|}
\hline\hline
   Source   & Period$^a$  &  $L_{\rm prop}(R)$$^b$               &  $\langle \dot{M}\rangle$$^c$ & $L_{\rm q}$$^d$  & $L_{\rm bb}$$^e$ &  $L_{\rm pl}$$^f$ \\
         &  (s) &  ($10^{33}$ erg s$^{-1}$)  & ($10^{-10}$ M$_{\odot}$ yr$^{-1}$) &  ($10^{33}$ erg s$^{-1}$) & ($10^{33}$ erg s$^{-1}$) & ($10^{33}$ erg s$^{-1}$) \\
\hline
4U 0115+63 & --  &  174 &  0.7 & 4.3 & 0.59$^{+0.07}_{-0.06}$ & 1.14$^{+0.20}_{-0.15}$ \\
V 0332+53 & --   &  991 &  1.0 & 5.5 & 0.26$^{+0.07}_{-0.06}$ & 0.54$^{+0.28}_{-0.16}$ \\
V 0332+53 (XMM) & --   &  991 & 1.0 & 5.5 & 0.34$^{+0.04}_{-0.04}$ & 0.59$^{+0.09}_{-0.08}$  \\
MXB 0656$-$072 &  --  &  0.3 &  0.2 & 1.3 & 3.80$^{+0.27}_{-0.25}$ & 5.62$^{+0.40}_{-0.38}$  \\
4U 0728$-$25 (XRT aver) & --    & $^g$ & 0.2 & 1.3 & 7.91$^{+0.57}_{-0.36}$ & 11.2$^{+0.5}_{-0.7}$ \\
RX J0812.4$-$3114 &  --  &  $^g$  & 0.4  & 2.2 & 1.60$^{+0.82}_{-0.62}$ & 2.54$^{+1.58}_{-1.05}$  \\
GS 0834$-$430 & --   &   $^g$  &  0.3 & 1.9 & 0.27$^{+0.29}_{-0.12}$  & 0.37$^{+0.27}_{-0.16}$ \\
GRO J1008$-$57 &  --  &  5.5 &  1.0  & 5.6 & 66.8$^{+1.6}_{-1.5}$ & 94.3$^{+2.2}_{-2.1}$  \\
2S 1417$-$624 & --   &   $^g$  &  0.3 & 2.1  & 2.01$^{+0.46}_{-0.38}$ &  2.66$^{+0.54}_{-0.40}$   \\
2S 1553$-$542 & --   &  119 & 1.5  & 8.8  & 1.05$^{+0.65}_{-0.44}$ & 2.24$^{+7.99}_{-1.09}$ \\
Swift J1626.6$-$5156 (10049) &   15.3360(6)  &  5.9 &   0.7 & 4.3 & 12.5$^{+0.6}_{-0.6}$ & 17.3$^{+0.8}_{-0.8}$  \\
Swift J1626.6$-$5156 (14642) &  --  &  5.9 &   0.7 & 4.3 & 2.87$^{+0.66}_{-0.48}$ & 4.15$^{+0.84}_{-0.70}$  \\
GS 1843+00 & --   &  5.1 & 1.2  & 7.3  & 1.55$^{+1.34}_{-0.66}$ & 2.00$^{+1.32}_{-0.77}$ \\
XTE J1946+274 &  15.760(3)  &  67 &   0.3  & 1.9 & 8.06$^{+0.78}_{-0.71}$ & 11.6$^{+1.1}_{-1.0}$ \\
KS 1947+300 &  18.802(3)  &  3.7 &   1.8  & 10.5  &  32.9$^{+2.4}_{-1.5}$ & 46.5$^{+3.3}_{-2.1}$ \\
SAX J21035+4545 &   350.7(1.1)   &   $^g$  &  0.5  & 3.2 & 1.49$^{+0.22}_{-0.16}$ &  2.42$^{+0.36}_{-0.31}$ \\
Cep X-4 &  66.38(3)  &  1.7 &  0.05 & 0.3 & 1.22$^{+0.15}_{-0.11}$  & 1.94$^{+0.19}_{-0.21}$ \\
SAX J2239.3+6116 & --   &  $^g$  &  0.06  & 0.4  & 1.01$^{+0.29}_{-0.23}$ & 1.46$^{+0.34}_{-0.27}$ \\
\hline
\end{tabular}
\begin{flushleft}{
$^{a}$Pulse period, if detected. Uncertainties are quoted at the $1\sigma$ level and calculated using the bootstrap approach \citep[for details see][]{2013AstL...39..375B}. \\
$^{b}$Propeller limiting  luminosity. \\
$^{c}$Long-term averaged mass accretion rate.\\
$^{d}$Expected quiescent  luminosity in the cooling scenario (using standard core cooling processes). \\
$^{e}$Unabsorbed luminosity in 0.5--10 keV band assuming a blackbody model. \\
$^{f}$Unabsorbed luminosity in 0.5--10 keV band assuming a power-law model. \\
$^g$Magnetic field is unknown.
  }\end{flushleft}
\end{table}
%\end{landscape}
%======================================================================================

 % Don't change these lines
\bsp    % typesetting comment
\label{lastpage}
\end{document}